\documentclass[aps,prb,twocolumn,floatfix]{revtex4-1}

\usepackage[utf8]{inputenc}
\usepackage{amsfonts}
\usepackage{amsmath}
\usepackage{amssymb}
\usepackage{bbm}
\usepackage{amsthm}
\usepackage{empheq}
\usepackage{enumitem}
\usepackage{verbatim}
\usepackage[dvips]{graphicx}
\usepackage[dvips]{color}
\usepackage{subfigure}

\usepackage{hyperref}

\begin{document}

\title{The dynamics of local magnetic moments induced by itinerant Weyl electrons}
\author{Predrag Nikoli\'c$^{1,2}$}
\affiliation{$^1$Department of Physics and Astronomy, George Mason University, Fairfax, VA 22030, USA}
\affiliation{$^2$Institute for Quantum Matter at Johns Hopkins University, Baltimore, MD 21218, USA}
\date{\today}

\begin{abstract}

We derive the effective interactions between local magnetic moments which are mediated by Weyl electrons in magnetic topological semimetals. The resulting spin dynamics is governed by the induced Heisenberg, Kitaev and Dzyaloshinskii-Moriya (DM) interactions with extended range and oscillatory dependence on the distance between the spins. These interactions are realized in multiple competing channels shaped by the multitude of Weyl nodes in the electron spectrum. Microscopic spins need to be spatially modulated with a channel-dependent wavevector in order to take advantage of the interactions. The DM vector is parallel to the displacement between the two interacting spins, and requires the presence of Weyl electron Fermi surfaces. We also derive the Weyl-induced chiral three-spin interaction in the presence of an external magnetic field. This interaction has an extended range as well, and acts upon the spatially modulated spins in various channels. Its tendency is to produce a skyrmion lattice or a chiral spin liquid which exhibits topological Hall effect. Ultimately, the theory developed here addresses magnetic dynamics in relativistic metals even when chiral magnetism is microscopically precluded. We discuss insights into the ordered state of the magnetic Weyl semimetal NdAlSi.

\end{abstract}

\maketitle

\section{Introduction}\label{secIntro}

Physical systems that combine strong interactions and non-trivial topology can exhibit many interesting phenomena. The most striking ones feature topological defects \cite{Mermin1979} -- either static in ordered phases such as the Abrikosov vortex lattice \cite{Abrikosov1957}, or delocalized in incompressible quantum liquids such as the fractional quantum Hall states \cite{Tsui1982}. Magnetic topological semimetals, where magnetism coexists with Dirac, Weyl or quadratic-band-touching electrons, are a new class of topological interacting materials envisioned theoretically \cite{Ari2010, Burkov2011a, Ran2011, Krempa2012} and gradually discovered experimentally \cite{Armitage2018}. Here, itinerant electrons with non-trivial band topology can develop magnetism themselves through a Fermi surface instability, or couple to an independent set of local magnetic moments. Regardless of whether the instability or intrinsic magnetism have any topological footprint, the itinerant electrons can transfer certain aspects of their topological dynamics to the magnetic moments. This process has been described in a general field theory framework \cite{Nikolic2019b}, but here we take a more concrete approach and reveal the experimentally relevant details of the Weyl electrons' influence on local moments.

In a broad sense, the following research is motivated by the quest for new chiral magnets and new magnetic states of matter. A Weyl spectrum of \emph{mobile} electrons can arise from a particular type of spin-orbit coupling. It has been shown recently \cite{Nikolic2019b} that the \emph{same} spin-orbit coupling acting on \emph{localized} electrons generates Dzyaloshinskii-Moriya (DM) and chiral spin interactions which may be able to introduce magnetic point-defects (hedgehogs) in the texture of the residual spins. A spin-orbit (i.e. DM) coupling \cite{Nagaosa2010b, Fritz2013, Chen2017b, Lounis2020, Hill2021}, and a chiral spin interaction enabled solely by an external magnetic field \cite{Chitra1995, Motrunich2006, Bulaevskii2008}, can also introduce magnetic line-defects (skyrmions). These are the direct microscopic origins of topological and chiral magnetism. We will show in this paper that itinerant Weyl electrons indirectly provide the same ingredients for the topological magnetism of local moments that they interact with. The observed manifestations of chiral magnetism include skyrmion \cite{Muhlbauer2009} and hedgehog \cite{Fujishiro2019} lattices, spin-momentum locking of spin waves \cite{Hoogdalem2013, Kovalev2014, RoldanMolina2016, Mook2017, Klinovaja2019}, spin wave topological bands \cite{Chen2016, Mook2016}, anomalous topological Hall effect \cite{Nagaosa2012, Hamamoto2015, Nakatsuji2016, Nakatsuji2017, Matsunoe2016, Wang2017, Ghimire2018, Parkin2016, Ohuchi2018}, and spin/thermal Hall effects \cite{Onose2010, Ong2015b, Okamoto2016, Zyuzin2016, Nakata2017, Mook2018}. Unconventional states such as chiral spin liquids are anticipated when quantum fluctuations delocalize the magnetic topological defects. The spin liquid variety associated with line defects has been possibly observed \cite{Machida2010, Balicas2011, Tokiwa2014}, while the proposed three-dimensional varieties  \cite{Cho2010, Maciejko2010, Fradkin2017, Nikolic2019} exhibit a fractional magnetoelectric effect and generalize the topological order of fractional quantum Hall states to higher dimensions. 


In a more concrete and experimentally-motivated \cite{Gaudet2021} context, this study explores the unique features of magnetism which are tied to the \emph{relativistic} nature of itinerant quasiparticles and their presence at multiple locations in the first Brillouin zone. Such features are somewhat independent of the Weyl electron chirality and can be experimentally observed even when the microscopic circumstances, such as spin anisotropy, preclude chiral magnetism.

The purpose of this paper is to analyze the s-d model of local moments coupled to itinerant Weyl electrons, and derive the electron-mediated interactions among the moments. The Weyl nodes are kept spherically symmetric in the present analysis, but otherwise form an arbitrary set whose total topological charge (chirality) adds up to zero in the first Brillouin zone. Every pair of Weyl nodes defines a separate channel for two-spin RKKY interactions. We find that the channels with equal-chirality nodes favor spatial modulations of the local magnetization with a period set by the difference between the node wavevectors. Therefore, it is easiest to describe the induced interactions in terms of ``rectified'' or ``staggered'' spins whose smooth ferromagnetic configuration represents the modulated microscopic spins. The two-spin (RKKY) interactions include Heisenberg, Kitaev and DM couplings between the ``rectified'' moments. We also obtain a three-spin chiral interaction in the presence of an external magnetic field. All couplings have an extended range (over a few lattice constants $a$), and additional algebraically-attenuated sign-changing oscillations in their dependence on the distance between spins. These features are controlled by the momentum cut-off $\Lambda$ of the linear Weyl spectrum, so the interactions appear short-ranged in the continuum limit $\pi a^{-1} > \Lambda \to \infty$.

Previous theoretical studies of RKKY interactions induced in Weyl semimetals \cite{JianhuiZhou2015, JianhuiZhou2017, Hosseini2015, Nomura2016} have also indicated the presence of Heisenberg and DM interactions, as well as the ``Ising'' interaction of the kind found in Kitaev models \cite{Kitaev2006b}. However, these studies do not agree in a number of important details, such as the orientation of vectors that characterize the DM and Kitaev interactions, and the spatial range of interactions. The present work attempts to resolve these discrepancies with a thorough and transparent calculation. We find agreement with Ref. \cite{JianhuiZhou2015, JianhuiZhou2017} in terms of the overall orientation of DM and Kitaev interactions, but the spatial range is different and compatible only with Ref.\cite{Nomura2016} which formulates its findings in the continuum limit. We also extend these earlier studies in several ways. The present analysis is not restricted to only two Weyl nodes, it goes beyond the two-spin interactions, and provides new relevant information for the modelling of spin dynamics in the experimentally explored magnetic Weyl semimetals \cite{Gaudet2021}. This paper also complements other related works \cite{Martin2008, Batista2014, Ozawa2016, Ozawa2017, Motome2017} by focusing on the effects specifically arising due to the spin-orbit coupling.

\subsection{The summary of results and outline}

Perhaps the most experimentally relevant finding of this study is that the pairs of equal-chirality Weyl nodes contribute \emph{ferromagnetic} Heisenberg interactions between two proximate ``rectified'' spins. This unfrustrated coupling is largest at short range and stimulates magnetic orders at wavevectors $\Delta{\bf Q} = {\bf Q}_m - {\bf Q}_n$ given by the locations ${\bf Q}_m, {\bf Q}_n$ of Weyl nodes in the first Brillouin zone. The uniform channel $\Delta{\bf Q}=0$ is made strongest by the contributions from every individual Weyl node through the intra-node scattering of electrons on local moments. The same-chirality inter-node scattering channels $\Delta{\bf Q}\neq 0$ can also be competitive, especially if $\Delta{\bf Q}$ is nearly commensurate with the lattice that the moments reside on. Spin modulations with incommensurate $\Delta{\bf Q}$ are possible, but disadvantaged at least at low temperatures when the magnetic order is to feature multiple wavevectors (e.g. the prominent $\Delta{\bf Q}=0$ and one or more $\Delta{\bf Q}\neq0$). This is due to the presence of higher modulation harmonics $n\Delta{\bf Q}$ ($n>1$), which are necessitated by the rigid magnitude of microscopic local spins but generally not favored by the locations of the Weyl nodes. All channels involving two opposite-chirality Weyl nodes are antiferromagnetic among the ``rectified'' spins. Their extended range can then introduce a geometric frustration for a dense arrangement of spins, so the ferromagnetic channels are naively expected to control the magnetic state.

This simple physical picture qualitatively explains the ordered state \cite{Gaudet2021} of the magnetic Weyl semimetal NdAlSi. Neutron scattering measurements have discovered a collinear easy-axis magnetic order of Nd moments, which combines a ferromagnetic component with spin modulations at the wavevector ${\bf q} = \left(\frac{2}{3}+\delta, \frac{2}{3}+\delta, 0\right)$ in the lattice constant units. The small incommensurate part $\delta$ corresponds to an amplitude modulation in an intermediate temperature range, and disappears below a lower critical temperature. At the same time, band structure calculations have identified a large number of Weyl nodes in both paramagnetic and ferromagnetic states near the Fermi level, unobscured by any sizeable conventional Fermi pocket. Some of these nodes form small chirality dipoles in momentum space, created by the spin-orbit coupling, with sets of dipoles separated by $\sim {\bf q}$ and its symmetry-related wavevectors. Other sets of Weyl nodes are found at different incommensurate separations. Given these spectral features, the observed magnetic order fits the naive expectation of ordering at both $\Delta{\bf Q}=0$ and $\Delta{\bf Q} \sim {\bf q}$ (the observed single-ion easy-axis anisotropy does not pose a critical obstacle to either channel). The incommensurate modulation by \emph{amplitude} is the result of fluctuations that resolve the frustration between the anisotropy, the desired smooth spin modulations at ${\bf q}$ and the microscopically rigid spin magnitude. When the temperature becomes too low, the fluctuations cannot soften the spins and the incommensurate component of the magnetic order becomes unsustainable.

The induced Kitaev and DM interactions are found to vanish at shortest distances, but still acquire the strength of the same order of magnitude as the Heisenberg coupling at finite distances. The ferromagnetic or antiferromagnetic nature of the Kitaev interaction (at distances where it is strongest) is the same as that of the Heisenberg coupling, so its main anticipated effect is to reduce the continuous magnetic symmetry down to a discrete group -- still supporting at least collinear orders as seen in NdAlSi \cite{Gaudet2021} (note that the crystal fields introduce spin anisotropy as well). More generally, the presence of sizable Kitaev interactions is interesting due to the prospects for stabilizing Kitaev spin liquids \cite{Kitaev2006b, Khaliullin2009, YongBaekKim2018, Nagler2019, Trivedi2019, YongBaekKim2020}. The Weyl-electron-induced DM interaction grows with the Fermi energy $|\mu|$ measured relative to the Weyl nodes, and vanishes when the nodes are exactly at the Fermi level. The DM vector ${\bf D} \propto {\bf r}_i-{\bf r}_j$ is parallel to the separation between the two interacting spins at locations ${\bf r}_i$ and ${\bf r}_j$. DM interactions generally support spin twists into ``spiral'' configurations, and this structure of ${\bf D}$ favors the emergence of skyrmions or hedgehogs in the spin texture \cite{Nikolic2019b}. However, any source of spin anisotropy goes against it and possibly leaves room only for a slight spin misalignment with the local easy axis directions. Such a misalignment is seen in NdAlSi, but has not been elucidated with sufficient detail yet \cite{Gaudet2021}.

The induced three-spin chiral interaction is perturbatively weaker than any two-spin interaction, and generally even more frustrated by the multitude of channels forged on the full set of Weyl nodes. Up to three nodes are involved in each channel, and the uniform channel contributed by each single node is the most significant. However, this interaction is proportional to the applied external magnetic field, and perhaps can be made strong in strong fields. Its main tendency is to stimulate skyrmions, magnetic line defects stretching in the direction of the field. Like its RKKY counterparts, this interaction features an algebraically attenuated oscillatory dependence of its coupling constant on the mutual separations between the spins (controlled by the momentum cut-off $\Lambda$). Due to the increasing complexity of calculations, we did not pursue four-spin and higher-order interactions. However, the four-spin interaction is potentially interesting since it provides an SU(2) part of the full U(1)$\times$SU(2) chiral interaction $(\phi + \hat{\bf n} \boldsymbol{\Phi})\, \hat{\bf n}_i(\hat{\bf n}_j\times\hat{\bf n}_k)$, where $\phi\propto B$ is the U(1) flux of the external magnetic field $B$ on the triangular plaquette formed by the spins $\hat{\bf n}_i, \hat{\bf n}_j, \hat{\bf n}_k$, and $\boldsymbol{\Phi}$ is the analogous SU(2) flux of the gauge field that captures the spin-orbit coupling. The SU(2) term is capable of stimulating skyrmions or hedgehogs without magnetic field depending on the type its non-Abelian flux. The analysis appropriate for localized electrons \cite{Nikolic2019b} suggests that Weyl electrons could favor the emergence of magnetic hedgehogs via this mechanism. Note that its perturbative weakness (at the 4th order) might also be compensated by a large strength of the spin-orbit coupling.

The paper layout is as follows. Section \ref{secEffH} introduces the effective model of Weyl electrons and local moments, and explains the general features of the perturbation theory that yields the effective interactions between the moments. The qualitative properties of the induced interactions, including their energy scales and scaling with the model parameters are deduced on general grounds before any calculations. Section \ref{sec2spin} proceeds with the technical derivation of the two-spin interactions, and presents the real-space properties of the Heisenberg, Kitaev and DM interactions at the end (section \ref{sec2spinRS}). Section \ref{sec3spin} derives the induced chiral three-spin interaction, and analyzes its real-space structure at the end (section \ref{sec3spinRS}). The final summary of conclusions and the discussion of theory limitations, extensions and applications is presented in Section \ref{secDiscussion}.

\raggedbottom

\section{Effective Hamiltonian of local moments}\label{secEffH}

Consider a simple model of local moments $\hat{{\bf n}}_{i}$ and itinerant electrons $\psi_{i}$ that live on a three-dimensional lattice with sites $i$:
\begin{equation}
H_{0}^{\phantom{\dagger}}=H_{n}^{\phantom{\dagger}}+\sum_{{\bf k}}\epsilon_{{\bf k}}^{\phantom{\dagger}}\psi_{{\bf k}}^{\dagger}\psi_{{\bf k}}^{\phantom{\dagger}}+J_{K}^{\phantom{\dagger}}\sum_i \hat{\bf n}_i^{\phantom{\dagger}}\,\psi_i^{\dagger}\boldsymbol{\sigma}\psi_i^{\phantom{\dagger}} \ .
\end{equation}
The moments and electrons interact via a Kondo or Hund coupling $J_{K}$. Both $\epsilon_{{\bf k}}$ and $J_{K}$ are energy scales, and the fields $\hat{\bf n}_i$ and $\psi_i$ are dimensionless. We will derive the effective Hamiltonian
\begin{equation}\label{Heff1}
H_{\textrm{eff}}^{\phantom{x}} = H_n + \sum_{n}^{\phantom{x}} \sum_{i_1\cdots i_n} J_{i_1\cdots i_n}^{a_1\cdots a_n} \, \hat{n}_{i_1}^{a_1} \cdots \hat{n}_{i_n}^{a_n}
\end{equation}
of local moments alone, which captures their emergent dynamics induced by the itinerant electrons. The intrinsic local moment dynamics ($H_n$) will be neglected assuming that the effective mass of the localized electrons is very large. We will use the units $\hbar=1$ and Einstein's convention of summation over the repeated spin-projection indices $a_j$.

\begin{figure}
\subfigure[{}]{\includegraphics[width=1.3in]{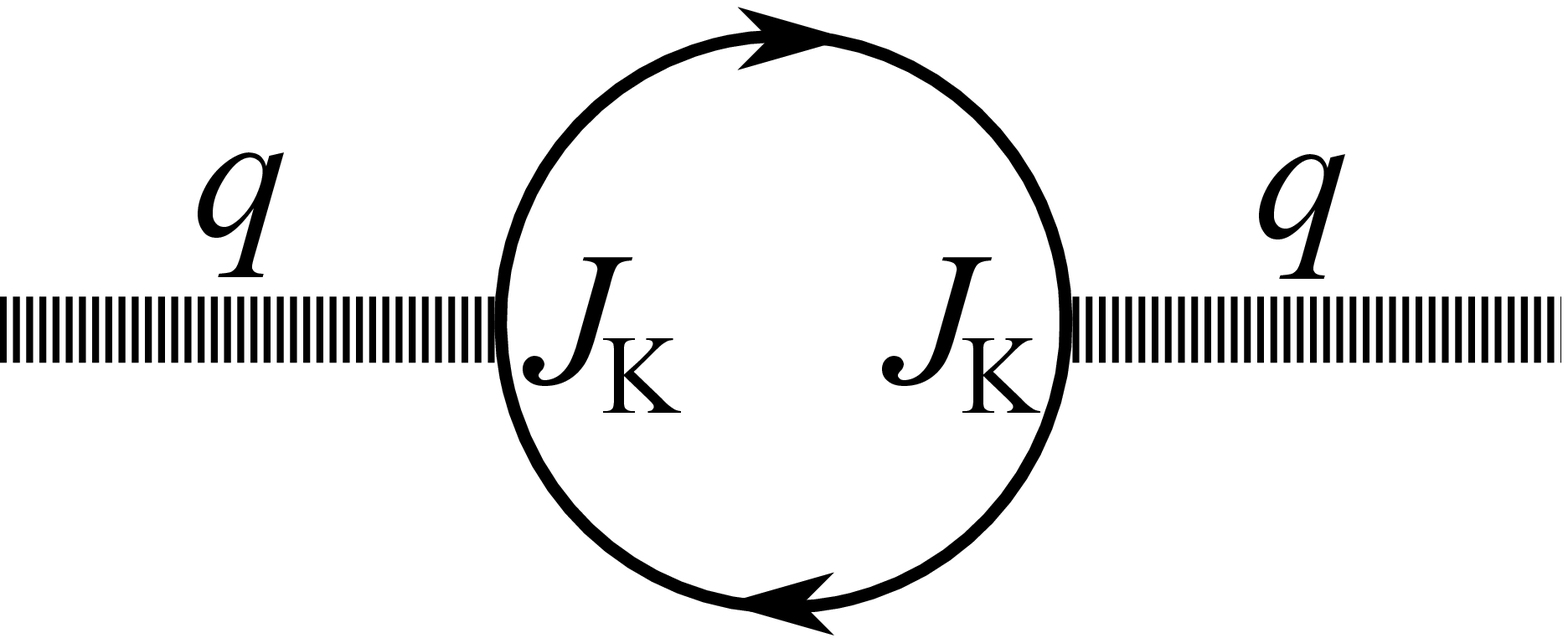}}
\subfigure[{}]{\includegraphics[width=1.3in]{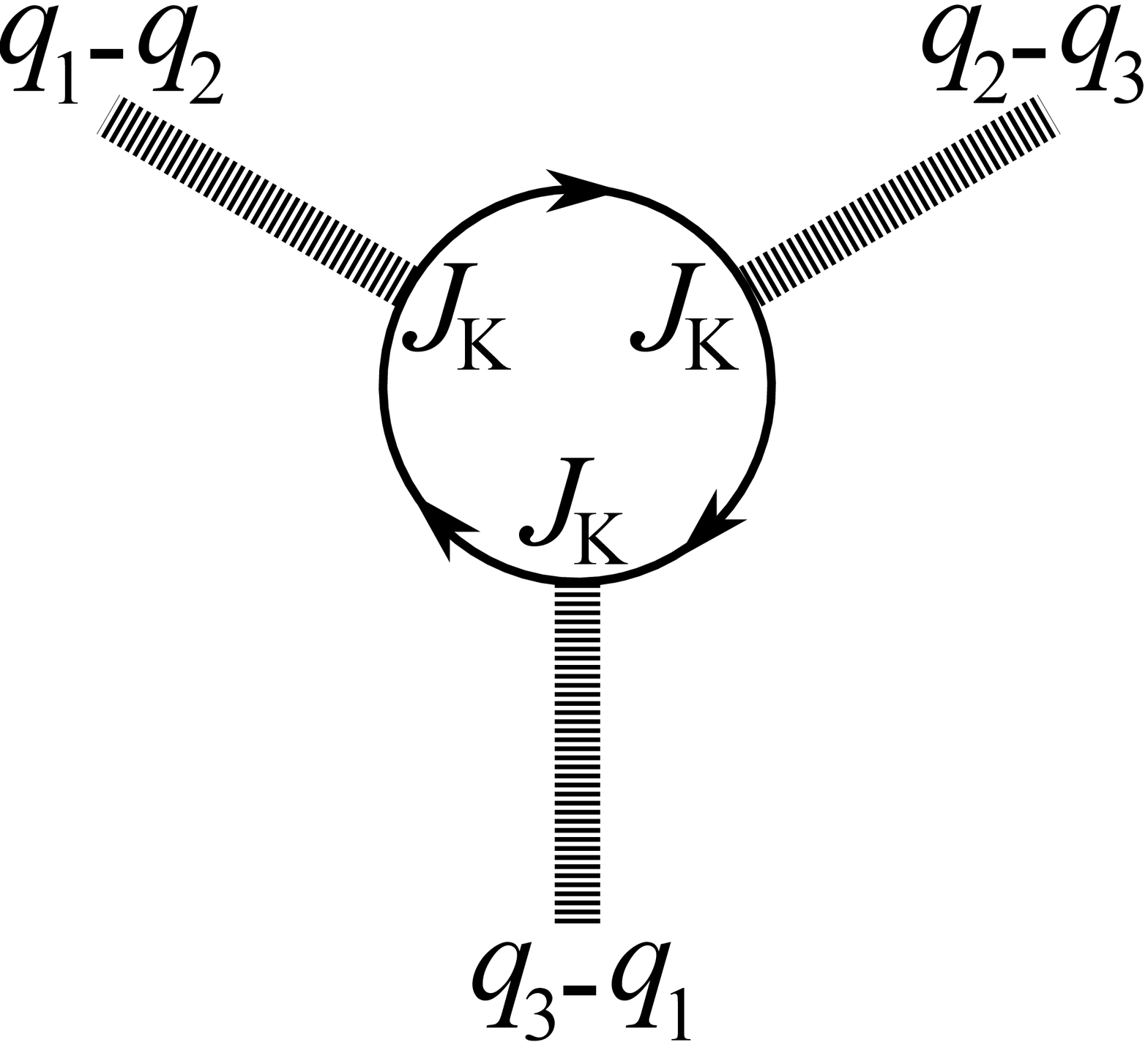}}
\caption{\label{Bubbles}Feynman diagrams for (a) two-spin and (b) three-spin interactions. Thick external lines represent local moment fields and thin lines represent Weyl electron propagators. The two-spin couplings will include Heisenberg, Kitaev and Dzyaloshinskii-Moriya interactions, and the three-spin coupling is the chiral interaction enabled by an external magnetic field or Weyl-node chirality in the presence of magnetic order.}
\end{figure}

The effective Hamiltonian $H_{\textrm{eff}}$ is extracted from the effective action $S_{\textrm{eff}}$ by integrating out the Grassmann spinor field $\psi$ in the continuum-limit path integral
\begin{equation}
e^{iS_{\textrm{eff}}[\hat{\bf n}_i]} \propto \int \mathcal{D}\psi\mathcal{D}\psi^\dagger\, e^{iS[\psi,\hat{\bf n}_i]}
\end{equation}
with real-time action
\begin{eqnarray}
S &=& \int\frac{d\omega}{2\pi}\frac{d^{3}k}{(2\pi)^{3}}\,\psi^{\dagger}(\omega,{\bf k})(\omega-\epsilon_{{\bf k}})\psi(\omega,{\bf k}) \\
&& -J_{K}\int\frac{d\omega}{2\pi}\frac{d\omega'}{2\pi}\frac{d^{3}k}{(2\pi)^{3}}\frac{d^{3}k'}{(2\pi)^{3}} \nonumber \\
&& \qquad\qquad \times \hat{{\bf n}}(\omega-\omega',{\bf k}-{\bf k}')\,\psi^{\dagger}(\omega',{\bf k}')\boldsymbol{\sigma}\psi(\omega,{\bf k}) \nonumber \ .
\end{eqnarray}
The local moments have been converted to continuum limit by the Fourier transform
\begin{equation}\label{nFourier}
\hat{\bf n}(\Omega,{\bf q})=a^3\sum_i\int dt\,e^{i({\bf q}{\bf r}_i-\Omega t)} \hat{\bf n}_i(t) \ ,
\end{equation}
where ${\bf r}_i$ are the discrete spatial coordinates of lattice sites and $a^3$ is the unit-cell volume (with a lattice constant $a$ on the cubic lattice). The perturbative expansion of the effective action
\begin{eqnarray}
S_{\textrm{eff}} &=& \sum_{n}\int\frac{d^{4}q_{1}}{(2\pi)^{4}}\cdots\frac{d^{4}q_{n}}{(2\pi)^{4}}\,(2\pi)^{4}\delta^4\left(\sum_{i=1}^{n}q_{i}\right) \\
&& \times \Gamma^{a_{1}\cdots a_{n}}(q_{1},\dots,q_{n})\,\hat{n}^{a_{1}}(q_{1})\cdots\hat{n}^{a_{n}}(q_{n}) \nonumber
\end{eqnarray}
is the sum of one-loop Feynman diagrams (Fig.\ref{Bubbles})
\begin{eqnarray}\label{Gamma1}
&& \Gamma^{a_{1}\cdots a_{n}}(q_{1},\dots,q_{n})=i\frac{J_{K}^{n}}{n}\int\frac{d^{4}k}{(2\pi)^{4}} \\
&& \quad \times \textrm{tr}\Bigl\lbrack G(k_1)\sigma^{a_{1}}G(k_2)\sigma^{a_{2}}\cdots G(k_n)\sigma^{a_{n}}\Bigr\rbrack \nonumber
\end{eqnarray}
with $k_m = k+\sum_{j=1}^{m-1}q_j$, because $\hat{\bf n}_i$ is not integrated out and all diagrams in the expansion are connected. Here, $q_i \equiv (\Omega_i, {\bf q}_i)$ combines the frequency and momentum in a single 4-vector, $\sigma^a$ are Pauli matrices, and the electron Green's functions $G(k)$ are spin matrices related to the time-ordered expectation values 
\begin{equation}\label{Green1}
\langle T\psi(k)\,\psi^{\dagger}(k')\rangle=i(2\pi)^{4}\delta^4(k-k')\,G(k) \nonumber
\end{equation}
in the non-interacting theory (with $J_{\textrm{K}}\to 0$). The $n^{\textrm{th}}$ order diagram $\Gamma$ involves $n$ external fields $\hat{\bf n}(q_i)$ and $n$ electron propagators, so it determines the coupling $J_{i_1\cdots i_n}^{a_1\cdots a_n} \propto J_{\textrm{K}}^n$ in (\ref{Heff1}). After the derivation of (\ref{Gamma1}), the real-space couplings in the effective Hamiltonian will be
\begin{eqnarray}\label{Jeff1}
J_{i_1\cdots i_n}^{a_1\cdots a_n} &=& -a^{3n} \int\frac{d^{3}q_{1}}{(2\pi)^{3}}\cdots\frac{d^{3}q_{n}}{(2\pi)^{3}}\,(2\pi)^{3}\delta^3\left(\sum_{i=1}^{n}{\bf q}_{i}\right) \nonumber \\
&& \times \Gamma^{a_{1}\cdots a_{n}}(q_{1},\dots,q_{n}) e^{i({\bf q}_1{\bf r}_{i_1}+\cdots+{\bf q}_{n}{\bf r}_{i_n})}
\end{eqnarray}
with $\Gamma$ evaluated at zero frequencies $\omega_1=\cdots=\omega_n=0$.

Our analysis will focus on an idealized magnetic Weyl semimetal whose electron spectrum $\epsilon_{\bf k}$ contains $N_{\textrm{W}}$ Weyl nodes with arbitrary chiralities $\chi_{n}=\pm 1$ at arbitrary wavevectors ${\bf Q}_{n}$, $n=1,\dots,N_{\textrm{W}}$. The periodic boundary conditions of the first Brillouin zone impose the requirement $\sum_n \chi_{n}=0$. We will assume for simplicity that all Weyl nodes sit at the same energy $\epsilon$=0 and have perfect spherical symmetry. Apart perhaps from the extreme dispersion tilting into type-II nodes, anisotropies of the Weyl spectrum are not expected to introduce substantial changes in the final results. The low-energy electrons associated with Weyl nodes must be formally handled using $N_{\textrm{W}}$ distinct Grassmann spinors $\psi_n$. These fields live at wavevectors ${\bf Q}_{n} + {\bf k}$ with ``small'' displacements $|{\bf k}|<\Lambda$ from the Weyl nodes at ${\bf Q}_{n}$. The momenta carried by local moments are restricted by momentum conservation to ${\bf Q}_{m}-{\bf Q}_{n}+{\bf q}$ with ``small'' ${\bf q}$. Therefore, the originally ``large'' values of $q$ on the left-hand-side in (\ref{Gamma1}) can select particular sets of Weyl nodes with fixed ``large'' parts ${\bf Q}_{m}-{\bf Q}_{n}$ of momentum transfers. This prompts us to switch to a more precise notation
\begin{eqnarray}\label{Gamma1b}
&& \Gamma_{m_1\cdots m_n}^{a_{1}\cdots a_{n}}(q_{1},\dots,q_{n})=i\frac{J_{K}^{n}}{n}\int\frac{d^{4}k}{(2\pi)^{4}} \\
&& \quad \times \textrm{tr}\Bigl\lbrack G_{m_1}(k_1)\sigma^{a_{1}}G_{m_2}(k_2)\sigma^{a_{2}}\cdots G_{m_n}(k_n)\sigma^{a_{n}}\Bigr\rbrack \nonumber
\end{eqnarray}
in which ${\bf q}_i, {\bf k}$ are understood to be the ``small'' wavevectors in the vicinity of the selected Weyl nodes at ${\bf Q}_m$. When we come back to the effective couplings on the lattice (\ref{Jeff1}), we will need to sum over all combinations of Weyl nodes $m_i$:
\begin{eqnarray}\label{Jeff1b}
&& J_{i_1\cdots i_n}^{a_1\cdots a_n} = -a^{3n} \sum_{\lbrace m\rbrace} e^{i\bigl(({\bf Q}_{m_2}-{\bf Q}_{m_1}){\bf r}_{i_1}+\cdots+({\bf Q}_{m_1}-{\bf Q}_{m_n}){\bf r}_{i_n}\bigr)} \nonumber \\
&& \qquad\times \int\frac{d^{3}q_{1}}{(2\pi)^{3}}\cdots\frac{d^{3}q_{n}}{(2\pi)^{3}}\,(2\pi)^{3}\delta^3\left(\sum_{i=1}^{n}{\bf q}_{i}\right) \nonumber \\
&& \qquad\times \Gamma_{m_1\cdots m_n}^{a_{1}\cdots a_{n}}(q_{1},\dots,q_{n}) e^{i({\bf q}_1{\bf r}_{i_1}+\cdots+{\bf q}_{m}{\bf r}_{i_n})}
\end{eqnarray}
Since ${\bf r}_i$ are the positions of local moments on lattice sites, the modulations at inter-node displacement wavevectors ${\bf Q}_m-{\bf Q}_n$ are formally evident. Multiple channels of such ``staggered'' modulations compete for their expression in the actual spin correlations or magnetic order. 

The Green's function of Weyl electrons in (\ref{Gamma1b}) is:
\begin{equation}\label{Green2}
G_{n}(\omega,{\bf k})=\ \Bigl\lbrack \omega-H_{n}({\bf k})+i\,\textrm{sign}(\epsilon_{n}({\bf k}))0^{+} \Bigr\rbrack^{-1} \ .
\end{equation}
The linear energy dispersion $\epsilon_{\bf k} \sim \pm v|{\bf k}|-\mu$, which extends only up to some cut-off momentum $|{\bf k}|<\Lambda$, can be captured by the simple Weyl Hamiltonian
\begin{equation}\label{WeylH}
H_{n}({\bf k})=v\chi_{n}\boldsymbol{\sigma}{\bf k}-\mu
\end{equation}
bundled with the chemical potential $\mu$. It will become apparent that the momentum integrals in the Feynman diagrams are ultra-violet divergent, so a required momentum cut-off $\Lambda$ will play a crucial role in providing a length scale even when the spectrum is formally relativistic ($\mu=0$). This will determine the spatial profile of the induced interactions with a resolution not better than $\Lambda^{-1} > ({\bf Q}_m-{\bf Q}_n)^{-1}$. When we substitute (\ref{Jeff1b}) into the effective Hamiltonian (\ref{Heff1}), we can define \emph{rectified} spins $\hat{\bf s}_i = \hat{\bf n}_i \exp\lbrack i({\bf Q}_m-{\bf Q}_n){\bf r}_i\rbrack$ by absorbing the modulation factors. The rectified spins can have smooth spatial variations on the lattice and represent entire clusters of the microscopic spins $\hat{\bf n}_i$ modulated at ${\bf Q}_m-{\bf Q}_n$. The effective Hamiltonian is then seen to capture the dynamics of rectified spins at length scales larger than $\Lambda^{-1}$.

The lattice length scale $a$ explicitly enters (\ref{Jeff1}) only as the shown overall factor, due to the final switch to real-space via (\ref{nFourier}). Then, the simplicity of the model admits only one relevant energy scale $v\Lambda$ at zero temperature and chemical potential. All couplings in the effective spin Hamiltonian can be expressed in the scaling form
\begin{equation}
J_{i_1\cdots i_n}^{a_1\cdots a_n} = v\Lambda \left(\frac{a^3 \Lambda^2 J_{\textrm{K}}}{v}\right)^n f^{a_1\cdots a_n} \left( \delta r_{ij}\,;\,\frac{\mu}{v\Lambda},\frac{B}{v\Lambda},\frac{T}{v\Lambda}\right)
\end{equation}
where $T$ is temperature in energy units, $B$ is magnetic field strength expressed as Zeeman energy, and $\delta r_{ij}$ are the relative lattice displacements between the spins in the interacting cluster. We will restrict the calculations to $T=0$, but the scaling form makes it apparent that low temperature is a small perturbation (the energy scale $v\Lambda$ is microscopic, hence easily larger than the room temperature $T\sim 25\;\textrm{meV}$).

\section{Two-spin interactions}\label{sec2spin}

Here we derive and analyze the two-spin interactions between local moments induced by the Weyl electrons. All such interactions arise from the bubble diagram in Fig.\ref{Bubbles}(a). We ought to calculate (\ref{Gamma1b}) at the second order of perturbation theory:
\begin{equation}\label{Gamma2b}
\Gamma^{ab}_{mn}(q) = i\frac{J_{K}^{2}}{2}\!\int\!\frac{d^{4}k}{(2\pi)^{4}} \textrm{tr}\left\lbrack G_{m}\!\left(k\!-\!\frac{q}{2}\right)\sigma^{a}G_{n}\!\left(k\!+\!\frac{q}{2}\right)\sigma^{b}\right\rbrack
\end{equation}
(note the shift of the integration variables). The ingredient of $\Gamma_{mn}^{ab}$ that we calculate first is the trace:
\begin{eqnarray}
&& \textrm{tr}\left\lbrack G_{m}\!\left(k\!-\!\frac{q}{2}\right)\sigma^{a}G_{n}\!\left(k\!+\!\frac{q}{2}\right)\sigma^{b}\right\rbrack = 
  2\,X^{ab}(\Omega,{\bf q};\omega,{\bf k})\, \prod_{s=\pm1} \nonumber \\[-0.1in]
&& ~~\times \frac{1}{\omega\!-\!\frac{\Omega}{2}\!\!+\!\mu\!-\!sv\chi_{m}\left|{\bf k}-\!\frac{\bf q}{2}\right|\!+\!i0^+\textrm{sign}\left(sv\chi_{m}\left|{\bf k}\!-\!\frac{\bf q}{2}\right|\!-\!\mu\right)} \nonumber \\
&& ~~\times \frac{1}{\omega\!+\!\frac{\Omega}{2}\!+\!\mu\!-\!sv\chi_{n}\left|{\bf k}\!+\!\frac{\bf q}{2}\right|\!+\!i0^+\textrm{sign}\left(sv\chi_{n}\left|{\bf k}\!+\!\frac{\bf q}{2}\right|\!-\!\mu\right)} \nonumber
\end{eqnarray}
where
\begin{eqnarray}
&& X^{ab}(\Omega,{\bf q};\omega,{\bf k}) = \left\lbrack (\omega+\mu)^{2}-\frac{\Omega^{2}}{4}\right\rbrack \delta^{ab} \\
&& ~~~~~ +v^{2}\chi_{m}\chi_{n}\left\lbrack 2\left(k^{a}k^{b}-\frac{q^{a}q^{b}}{4}\right)-\delta^{ab}\left(|{\bf k}|^2-\frac{|{\bf q}|^2}{4}\right)\right\rbrack \nonumber \\
&& ~~~~~ +iv\epsilon^{abc}\biggl\lbrack \chi_{m}\left(\omega+\mu+\frac{\Omega}{2}\right)\left(k^{c}-\frac{q^{c}}{2}\right) \nonumber \\
&& ~~\qquad\quad\quad -\chi_{n}\left(\omega+\mu-\frac{\Omega}{2}\right)\left(k^{c}+\frac{q^{c}}{2}\right)\biggr\rbrack \nonumber \ .
\end{eqnarray}
We used (\ref{Green2}), (\ref{WeylH}) together with $\textrm{tr}(1)=2$, $\textrm{tr}(\sigma^{a})=0$, and $\sigma^{a}\sigma^{b}=\delta^{ab}+i\epsilon^{abc}\sigma^{c}$ involving the Levi-Civita tensor $\epsilon^{abc}$. Frequency $\omega$ integration in (\ref{Gamma2b}) is straight-forward:
\begin{widetext}
\begin{eqnarray}
&& \Gamma^{ab}_{mn}(q) = -\frac{J_{K}^{2}}{2}\sum_{s=\pm1}\int\frac{d^{3}k}{(2\pi)^{3}}
\Biggl\lbrace \frac{sX^{ab}\left(\Omega,{\bf q};sv\chi_{m}\left|{\bf k}-\frac{\bf q}{2}\right|+\frac{\Omega}{2}-\mu,{\bf k}\right)\,\theta\left(-sv\chi_{m}\left|{\bf k}-\frac{\bf q}{2}\right|+\mu\right)}{v\chi_{m}\left|{\bf k}-\frac{\bf q}{2}\right|\,\left\lbrack\left(sv\chi_{m}\left|{\bf k}-\frac{\bf q}{2}\right|+\Omega\right)^{2}-v^{2}\left|{\bf k}+\frac{\bf q}{2}\right|^{2}\right\rbrack} \\
&& \qquad\qquad\qquad\qquad\qquad\qquad\quad +\frac{sX^{ab}\left(\Omega,{\bf q};sv\chi_{n}\left|{\bf k}+\frac{\bf q}{2}\right|-\frac{\Omega}{2}-\mu,{\bf k}\right)\,\theta\left(-sv\chi_{n}\left|{\bf k}+\frac{\bf q}{2}\right|+\mu\right)}{v\chi_{n}\left|{\bf k}+\frac{\bf q}{2}\right|\,\left\lbrack\left(sv\chi_{n}\left|{\bf k}+\frac{\bf q}{2}\right|-\Omega\right)^{2}-v^{2}\left|{\bf k}-\frac{\bf q}{2}\right|^{2}\right\rbrack} \Biggr\rbrace \nonumber \ .
\end{eqnarray}
\end{widetext}
Here, $\theta(x)$ is the step function. We neglected the residual imaginary infinitesimal terms in the denominators since we are currently not interested in dissipative processes.

Since $\Gamma_{mn}^{ab}(\Omega,{\bf q})$ is analytic at $\Omega=0$ for generic non-zero ${\bf q}$, we can expand it in powers of $\Omega$ and associate the zeroth-order term in the expansion with the instantaneous interactions between the spins. Therefore, taking the limit $\Omega\to0$ provides access to the non-retarded part of the effective spin interactions and enables further simplifications:
\begin{widetext}
\begin{eqnarray}\label{Gamma2c}
&& \Gamma^{ab}_{mn}({\bf q}) = \frac{J_{\textrm{K}}^2}{8v^3} \int\frac{d^3 k}{(2\pi)^3} \frac{1}{\bf kq} \Biggl\lbrack
\frac{X_{-}^{ab}-\left(X_{-}^{ab}\right)^{*}}{\chi_{m}\left\vert {\bf k}-\frac{{\bf q}}{2}\right\vert}
-\frac{X_{-}^{ab}\,\textrm{sign}\left(v\chi_{m}\left\vert {\bf k}-\frac{{\bf q}}{2}\right\vert\!-\mu\right)+\left(X_{-}^{ab}\right)^{*}\textrm{sign}\left(v\chi_{m}\left\vert {\bf k}-\frac{{\bf q}}{2}\right\vert+\mu\right)}{\chi_{m}\left\vert {\bf k}-\frac{{\bf q}}{2}\right\vert}\qquad \\
&& \qquad\qquad\qquad\qquad\qquad\quad
-\frac{X_{+}^{ab}-\left(X_{+}^{ab}\right)^{*}}{\chi_{n}\left\vert {\bf k}+\frac{{\bf q}}{2}\right\vert}
+\frac{X_{+}^{ab}\,\textrm{sign}\left(v\chi_{n}\left\vert {\bf k}+\frac{{\bf q}}{2}\right\vert-\mu\right)+\left(X_{+}^{ab}\right)^{*}\!\textrm{sign}\left(v\chi_{n}\left\vert {\bf k}+\frac{{\bf q}}{2}\right\vert+\mu\right)}{\chi_{n}\left\vert {\bf k}+\frac{{\bf q}}{2}\right\vert} \Biggr\rbrack \nonumber
\end{eqnarray}
with
\begin{equation}
X_{\pm}^{ab} = v^{2}\left\vert {\bf k}\!\pm\!\frac{\bf q}{2}\right\vert^{2}\delta^{ab}
  +v^{2}\chi_{m}\chi_{n}\left( 2k^{a}k^{b}-\delta^{ab}|{\bf k}|^2
    -\frac{2q^{a}q^{b}\!-\!\delta^{ab}|{\bf q}|^2}{4}\right)
  -iv^{2}\epsilon^{abc}\left\vert{\bf k}\!\pm\!\frac{\bf q}{2}\right\vert
    \left\lbrack (1\!+\!\chi_{m}\chi_{n})\frac{q^{c}}{2}\pm(1\!-\!\chi_{m}\chi_{n})k^{c}
    \right\rbrack \ . \nonumber
\end{equation}
\end{widetext}
Note that the terms without sign functions in (\ref{Gamma2c}) vanish in typical perturbative treatments of Fermi liquids by the virtue of $X\in\mathbb{R}$. Here, however, $X\in\mathbb{C}$ originates in the chirality of Weyl nodes and yields Dzyaloshinskii-Moriya interactions between the local moments.

Now we integrate out the momentum ${\bf k}$ using a polar coordinate system whose $z$-axis is aligned with ${\bf q}$. Choosing the polar instead of e.g. the spherical coordinate system affects only the manner in which we capture the contributions of high-energy states, since we must introduce a momentum cut-off $\Lambda$. However, this theory is anyway invalid at high energies, so its value is the ability to reveal the universal low-energy features of spin dynamics instead of the numerically accurate details that depend on the microscopic properties of electrons on the lattice. Writing ${\bf k} = k_\parallel\hat{\bf z} + {\bf k}_\perp = (k_{\parallel},k_{\perp},\theta)$ and ${\bf q} = q\hat{\bf z}$ we have
\begin{equation}
{\bf k}{\bf q}=k_{\parallel}^{\phantom{x}}q\quad,\quad\left\vert {\bf k}\pm\frac{{\bf q}}{2}\right\vert =\sqrt{\left(k_{\parallel}^{\phantom{x}}\pm\frac{q}{2}\right)^{2}+k_{\perp}^{2}} \ .
\end{equation}
The polar coordinate system is convenient because these expressions do not depend on the polar angle $\theta$. The only quantities that depend on $\theta$ are the parts of $X_\pm^{ab}$ which are linear or quadratic in the components of ${\bf k}_\perp$; they either integrate out to zero or average out to a half of the maximum value, while all other parts pick a factor of $2\pi$ from the angle integration. Hence, we get:
\begin{widetext}
\begin{eqnarray}\label{Gamma2d}
\Gamma^{ab}_{mn}({\bf q}) &=& \frac{J_{K}^{2}}{8(2\pi)^{2}vq}\int\limits _{-\infty}^{\infty}dk_{\parallel}\int\limits_{0}^{\infty}dk_{\perp}\,\frac{k_{\perp}^{\phantom{x}}}{k_{\parallel}^{\phantom{x}}}\Biggl\lbrack \chi_{m}^{\phantom{x}}\Bigl(Y_{-}^{ab}-\left(Y_{-}^{ab}\right)^{*}\Bigr)-\chi_{n}^{\phantom{x}}\Bigl(Y_{+}^{ab}-\left(Y_{+}^{ab}\right)^{*}\Bigr) \\
&& \qquad\qquad\qquad -Y_{-}^{ab}\,\textrm{sign}\left(v\sqrt{\left(k_{\parallel}^{\phantom{x}}-\frac{q}{2}\right)^{2}+k_{\perp}^{2}}-\chi_{m}\mu\right)-\left(Y_{-}^{ab}\right)^{*}\!\textrm{sign}\left(v\sqrt{\left(k_{\parallel}^{\phantom{x}}-\frac{q}{2}\right)^{2}+k_{\perp}^{2}}+\chi_{m}\mu\right)\nonumber \\
&& \qquad\qquad\qquad +Y_{+}^{ab}\,\textrm{sign}\left(v\sqrt{\left(k_{\parallel}^{\phantom{x}}+\frac{q}{2}\right)^{2}+k_{\perp}^{2}}-\chi_{n}\mu\right)+\left(Y_{+}^{ab}\right)^{*}\!\textrm{sign}\left(v\sqrt{\left(k_{\parallel}^{\phantom{x}}+\frac{q}{2}\right)^{2}+k_{\perp}^{2}}+\chi_{n}\mu\right)\Biggr\rbrack \nonumber
\end{eqnarray}
\end{widetext}
where we defined
\begin{equation}\label{Y0}
Y_{\pm}^{ab}=\frac{X_{\pm}^{ab}}{v^2\sqrt{\left(k_{\parallel}^{\phantom{x}}\pm\frac{q}{2}\right)^{2}+k_{\perp}^{2}}} \ .
\end{equation}
Form this point on, we will separately consider the pairs of Weyl nodes with the same $\chi_{m}\chi_{n}=+1$ and opposite $\chi_{m}\chi_{n}=-1$ chiralities. We will also specialize to concrete values of spin indices $a,b\in\lbrace\parallel,\perp,\perp'\rbrace$, where $\parallel$ indicates the direction along ${\bf q}$ and $\perp,\perp'$ are any two orthogonal directions both perpendicular to ${\bf q}$.

\subsubsection{Same-chirality nodes}

With $\chi_{m}=\chi_{n}$, the formulas (\ref{Y0}) reduce to:
\begin{eqnarray}\label{Y1}
Y_{\pm}^{\parallel\parallel}=\frac{2k_{\parallel}^{\phantom{x}}\left(k_{\parallel}^{\phantom{x}}\!\pm\!\frac{q}{2}\right)}{\sqrt{\left(k_{\parallel}^{\phantom{x}}\!\pm\!\frac{q}{2}\right)^{2}\!\!+\!k_{\perp}^{2}}}&\quad,\quad& Y_{\pm}^{\perp\perp}=\frac{k_{\perp}^{2}\!\pm\! q\left(k_{\parallel}^{\phantom{x}}\!\pm\!\frac{q}{2}\right)}{\sqrt{\left(k_{\parallel}^{\phantom{x}}\!\pm\!\frac{q}{2}\right)^{2}\!\!+\!k_{\perp}^{2}}} \nonumber \\[0.05in]
Y_{\pm}^{\perp\perp'}=-iq\,\epsilon^{\parallel\perp\perp'} &\quad,\quad& Y_{\pm}^{\parallel\perp}=Y_{\pm}^{\perp\parallel}=0
\end{eqnarray}
It is convenient to carry out the remaining integrations using dimensionless variables
\begin{equation}
\zeta=\frac{vq}{2|\mu|}\quad,\quad\xi=\frac{vk_{\parallel}}{|\mu|}\quad,\quad\eta=\frac{vk_{\perp}}{|\mu|} \ .
\end{equation}
Observing that
\begin{equation}
\textrm{sign}\left(u\!-\!\chi_{m}\mu\right)-\textrm{sign}\left(u\!+\!\chi_{m}\mu\right)=-2\chi_{m}\,\textrm{sign}(\mu)\:\theta\left(|\mu|\!-\!u\right) \nonumber
\end{equation}
holds for $u>0$, the calculation of the ``chiral'' two-spin interaction is straight-forward:
\begin{eqnarray}\label{Gamma2e0}
&& \Gamma_{mn}^{\perp\perp'}({\bf q}) = -i\,\epsilon^{\parallel\perp\perp'}\, \frac{\chi_{m}\,\textrm{sign}(\mu)\,J_{K}^{2}\mu^{2}}{4(2\pi)^{2}v^{3}} \!\int\limits_{-\infty}^{\infty}
  \!\!d\xi\int\limits_{0}^{\infty}\!d\eta \\
&& \quad\; \times \frac{\eta}{\xi} \left\lbrack \theta\left(1\!-\!\sqrt{\left(\xi\!-\!\zeta\right)^{2}\!+\!\eta^{2}}\right)-\theta\left(1\!-\!\sqrt{\left(\xi\!+\!\zeta\right)^{2}\!+\!\eta^{2}}\right)\right\rbrack \nonumber \\
&& \quad = i\,\epsilon^{\parallel\perp\perp'}\,\frac{\chi_{m}\,\textrm{sign}(\mu)\,J_{K}^{2}\mu^{2}}{4(2\pi)^{2}v^{3}}\left\lbrack (1-\zeta^{2})\log\left\vert \frac{1-\zeta}{1+\zeta}\right\vert -2\zeta\right\rbrack \nonumber
\end{eqnarray}
Note that the dissipative terms were discarded by taking the principal value of the integral.

The remaining non-zero components of the $\Gamma_{mn}$ tensor are diagonal and involve the real-valued $Y_{\pm}^{\parallel\parallel}$ and $Y_{\pm}^{\perp\perp}$. Observing that
\begin{equation}
\textrm{sign}\left(u\!-\!\chi_{m}\mu\right)+\textrm{sign}\left(u\!+\!\chi_{m}\mu\right) = 2\,\theta\left(u\!-\!|\mu|\right) \nonumber
\end{equation}
holds for $u>0$, we find that (\ref{Gamma2d}) with $a=b$ becomes:
\begin{equation}\label{Gamma2e}
\Gamma_{mn}^{ab}({\bf q}) = -\frac{J_{K}^{2}\mu^{2}}{16(2\pi)^{2}v^{3}\zeta}\int\limits _{-\infty}^{\infty}dx\!\int\limits _{1-x^{2}}^{\infty}\!\!\!d(\eta^{2})\left(\frac{\bar{Y}_{-}^{ab}}{x+\zeta}-\frac{\bar{Y}_{+}^{ab}}{x-\zeta}\right)
\end{equation}
Here, $x=\xi\pm\zeta$ is a shifted integration variable, and the factors (\ref{Y1}) have been made dimensionless:
\begin{equation}
\bar{Y}_{\pm}^{\parallel\parallel}=\frac{2(x\mp\zeta)x}{\sqrt{x^{2}+\eta^{2}}}\quad,\quad\bar{Y}_{\pm}^{\perp\perp}=\frac{\eta^{2}\pm2\zeta x}{\sqrt{x^{2}+\eta^{2}}} \ .
\end{equation}
In order to proceed, we need two integrals:
\begin{eqnarray}
I_1 &=& \int\limits _{-\infty}^{\infty}dx\,\left(\frac{1}{x+\zeta}-\frac{1}{x-\zeta}\right)\!\int\limits_{1-x^{2}}^{\infty}\!\!\!\frac{d(\eta^{2})\,\eta^{2}}{\sqrt{x^{2}+\eta^{2}}} \\
&\to& 2\int\limits _{-\infty}^{\infty}dx\left(\frac{1}{x+\zeta}-\frac{1}{x-\zeta}\right) \nonumber \\&&\quad \times\left\lbrack \lambda^{2}\sqrt{\lambda^{2}\!+\!x^{2}}-\frac{2}{3}(\lambda^{2}\!+\!x^{2})^{3/2}-(1\!-\!x^{2})+\frac{2}{3}\right\rbrack \nonumber \\
&\to& -8\zeta\lambda^{2}\int\limits_{0}^{1}dy\,\frac{\sqrt{1+y^{2}}-\frac{2}{3}(1+y^{2})^{3/2}}{y^{2}-(\zeta/\lambda)^{2}} + \cdots \nonumber \\
&=& -8\zeta\Bigl(-0.942809\lambda^{2}+0.040957\zeta^{2}+\cdots\Bigr) \nonumber \ ,
\end{eqnarray}
and
\begin{eqnarray}
I_2 &=& \int\limits _{-\infty}^{\infty}dx\,\left(\frac{1}{x+\zeta}+\frac{1}{x-\zeta}\right)x\!\int\limits_{1-x^{2}}^{\infty}\!\!\!\frac{d(\eta^2)}{\sqrt{x^{2}+\eta^2}} \\
&\to& 2\int\limits _{-\infty}^{\infty}dx\,\left(\frac{1}{x+\zeta}+\frac{1}{x-\zeta}\right)x\,\Bigl(\sqrt{x^{2}+\lambda^{2}}-1\Bigr) \nonumber \\
&\to& 8\lambda^{2}\int\limits_{0}^{1}dy\,\frac{y^{2}\sqrt{1+y^{2}}}{y^{2}-(\zeta/\lambda)^{2}} + \cdots \nonumber \\
&=& 8\Bigl(1.14779\lambda^{2}-0.53284\zeta^{2}+\cdots\Bigr) \nonumber \ .
\end{eqnarray}
Both integrals are divergent and require an ultra-violet cut-off $\lambda=v\Lambda/|\mu|$. Arrows represent replacements of the ultra-violet integral bounds with $\lambda$, and the final results show only the leading powers of $\lambda$ for each power of $\zeta$ (i.e. the numerical coefficients in front of $\lambda^2, \zeta^2$ are the lowest-order terms in the respective expansions over $\lambda^{-1}=|\mu|/v\Lambda$). We readily find
\begin{eqnarray}\label{Gamma2e1}
\Gamma_{mn}^{\parallel\parallel}({\bf q}) &=& 0 \\
\Gamma_{mn}^{\perp\perp}({\bf q}) &=& -\frac{J_{K}^{2}\mu^{2}}{16(2\pi)^{2}v^{3}\zeta}(I_{1}-2\zeta I_{2}) \nonumber \\
  &=& -\frac{J_{K}^{2}}{2(2\pi)^{2}v}\Bigl(-1.352771\,\Lambda^{2}+0.25618075\,q^{2}\Bigr) \nonumber
\end{eqnarray}
to the leading order in $\Lambda$ and $q$.

\subsubsection{Opposite-chirality nodes}

With $\chi_{m}=-\chi_{n}$, the formulas (\ref{Y0}) reduce to:
\begin{eqnarray}\label{Y2}
Y_{\pm}^{\parallel\parallel}=\frac{2k_{\perp}^{2}\!\pm\! q\left(k_{\parallel}^{\phantom{x}}\!\pm\!\frac{q}{2}\right)}{\sqrt{\left(k_{\parallel}^{\phantom{x}}\!\pm\!\frac{q}{2}\right)^{2}\!\!+\!k_{\perp}^{2}}} &\quad,\quad& Y_{\pm}^{\perp\perp}=\frac{k_{\perp}^{2}\!+\!2k_{\parallel}^{\phantom{x}}\left(k_{\parallel}^{\phantom{x}}\!\pm\!\frac{q}{2}\right)}{\sqrt{\left(k_{\parallel}^{\phantom{x}}\!\pm\!\frac{q}{2}\right)^{2}\!\!+\!k_{\perp}^{2}}} \nonumber \\
Y_{\pm}^{\perp\perp'}=\mp 2ik_{\parallel}^{\phantom{x}}\,\epsilon^{\parallel\perp\perp'} &\quad,\quad& Y_{\pm}^{\parallel\perp}=Y_{\pm}^{\perp\parallel}=0
\end{eqnarray}
It is easy to see that the features $Y_{-}^{\perp\perp'}=-Y_{+}^{\perp\perp'}$ and $\chi_{m}=-\chi_{n}$ cause the cancellation of all contributions to the ``chiral'' two-spin interactions:
\begin{equation}
\Gamma_{mn}^{\perp\perp'} = 0 \ .
\end{equation}
The physical implication is that the electron scattering between opposite-chirality Weyl nodes does not contribute to the Dzyaloshinskii-Moriya (DM) interactions between local moments. The DM interaction is an SU(2) gauge field coupled to spin currents, which in this case takes the form $A_{i}^{a}\propto\delta_{i}^{a}$ that violates the inversion symmetry and respects the time-reversal (TR) symmetry. If it emerged from the opposite-chirality nodes, it could arise in a TR-breaking Weyl semimetal with only two nodes. In such a simple Weyl semimetal, the intra-node (same-chirality) channel does not produce a DM interaction either because each Weyl node yields a uniform $A_{i}^{a}\propto\chi_{i}$ and $\sum\chi_{i}=0$ in the first Brillouin zone. This is, of course, consistent with the presence of inversion symmetry.

The other non-zero components of the $\Gamma_{mn}$ tensor are determined using the same procedure as in the previous section. The formula (\ref{Gamma2e}) applies for the opposite-chirality nodes as well, provided that we use the appropriate dimensionless versions of (\ref{Y2}):
\begin{equation}
\bar{Y}_{\pm}^{\parallel\parallel}=2\frac{\eta^{2}\pm\zeta x}{\sqrt{x^{2}+\eta^{2}}}\quad,\quad\bar{Y}_{\pm}^{\perp\perp}=\frac{\eta^{2}+2(x\mp\zeta)x}{\sqrt{x^{2}+\eta^{2}}}
\end{equation}
(recall $x=\xi\pm\zeta$). Then:
\begin{eqnarray}\label{Gamma2e2}
\Gamma_{mn}^{\parallel\parallel}({\bf q}) &=& -\frac{J_{K}^{2}\mu^{2}}{8(2\pi)^{2}v^{3}\zeta}(I_{1}-\zeta I_{2}) \\
&=& -\frac{J_{K}^{2}}{2(2\pi)^{2}v}\Bigl(-0.409962\,\Lambda^{2}+0.2459415\,q^{2}\Bigr) \nonumber \\
\Gamma_{mn}^{\perp\perp}({\bf q}) &=& -\frac{J_{K}^{2}\mu^{2}}{16(2\pi)^{2}v^{3}\zeta}\,I_{1} \nonumber \\
&=& -\frac{J_{K}^{2}}{2(2\pi)^{2}v}\Bigl(0.942809\,\Lambda^{2}-0.01023925\,q^{2}\Bigr) \ . \nonumber
\end{eqnarray}

\subsection{Induced interactions in real space}\label{sec2spinRS}

Here we determine the interaction coupling $J_{ij}^{ab}$ in the effective spin Hamiltonian (\ref{Heff1})
\begin{equation}\label{Heff2}
H_{\textrm{eff}}^{\phantom{x}} = \sum_{ij} J_{ij}^{ab} \, \hat{n}_{i}^{a} \hat{n}_{j}^{b} + \cdots
\end{equation}
using (\ref{Jeff1b}):
\begin{equation}\label{Jeff2}
J_{ij}^{ab} = -a^{6} \sum_{m,n} e^{i({\bf Q}_{n}-{\bf Q}_{m})({\bf r}_{i}-{\bf r}_{j})} \!\int\!\frac{d^{3}q}{(2\pi)^{3}}\, \Gamma_{mn}^{ab}({\bf q})\, e^{i{\bf q}({\bf r}_{i}-{\bf r}_{j})}
\end{equation}
It will be convenient to express the contribution of any Weyl node pair $m,n$ to (\ref{Heff2}) as the interaction between complex ``rectified'' spins
\begin{equation}\label{RectSpin}
\hat{\bf s}_i = e^{i({\bf Q}_{n}-{\bf Q}_{m}){\bf r}_{i}} \hat{\bf n}_i \ .
\end{equation}
Microscopic spins $\hat{\bf n}_i$ will tend to form clusters modulated at ``large'' wavevectors ${\bf Q}_n-{\bf Q}_m$, and these clusters will be represented by the ``smooth'' field $\hat{\bf s}_i$.

$\Gamma_{mn}^{ab}$ given by (\ref{Gamma2e0}), (\ref{Gamma2e1}) and (\ref{Gamma2e2}) was originally constructed in the basis $(\hat{\bf q}, \hat{\boldsymbol{\theta}}, \hat{\boldsymbol{\phi}})$ for spin vectors. We will convert $\Gamma_{mn}^{ab}$ to the fixed basis $(\hat{\bf x}, \hat{\bf y}, \hat{\bf z})$ aligned with ${\bf r}_{ij} = {\bf r}_i - {\bf r}_j = |{\bf r}_{ij}| \hat{\bf z}$ and solve the integral in (\ref{Jeff2}) using spherical coordinates ${\bf q}=(q,\theta,\phi)$. The matrix representation of $\Gamma_{mn}^{ab}$ in the fixed basis is
\begin{equation}\label{Gamma2f}
\hat{\Gamma}_{mn}^{\phantom{x}} = \hat{M}_\parallel^{\phantom{x}} \Gamma_{mn}^{\parallel\parallel} + \hat{M}_\perp^{\phantom{x}} \Gamma_{mn}^{\perp\perp} + \hat{M}_{\textrm{ch}}^{\phantom{x}} \Gamma_{mn}^{\perp\perp'}
\end{equation}
where
\begin{equation}
\hat{M}_\parallel^{\phantom{x}} = \left(\begin{array}{ccc}
\cos^{2}\phi\sin^{2}\theta & \frac{\sin(2\phi)}{2}\sin^{2}\theta & \cos\phi\frac{\sin(2\theta)}{2}\\
\frac{\sin(2\phi)}{2}\sin^{2}\theta & \sin^{2}\phi\sin^{2}\theta & \sin\phi\frac{\sin(2\theta)}{2}\\
\cos\phi\frac{\sin(2\theta)}{2} & \sin\phi\frac{\sin(2\theta)}{2} & \cos^{2}\theta
\end{array}\right) \nonumber
\end{equation}
\begin{equation}
\hat{M}_\perp^{\phantom{x}} = \left(\begin{array}{ccc}
1\!-\!\cos^{2}\phi\sin^{2}\theta & -\frac{\sin(2\phi)}{2}\sin^{2}\theta & -\!\cos\phi\frac{\sin(2\theta)}{2}\\
-\frac{\sin(2\phi)}{2}\sin^{2}\theta & 1\!-\!\sin^{2}\phi\sin^{2}\theta & -\!\sin\phi\frac{\sin(2\theta)}{2}\\
-\!\cos\phi\frac{\sin(2\theta)}{2} & -\!\sin\phi\frac{\sin(2\theta)}{2} & \sin^{2}\theta
\end{array}\right) \nonumber
\end{equation}
\begin{equation}
\hat{M}_{\textrm{ch}}^{\phantom{x}} = \left(\begin{array}{ccc}
0 & \cos\theta & -\sin\phi\sin\theta\\
-\cos\theta & 0 & \cos\phi\sin\theta\\
\sin\phi\sin\theta & -\cos\phi\sin\theta & 0
\end{array}\right) \nonumber
\end{equation}
It becomes quickly apparent that integrating out $\phi$ preserves only those matrix elements of $\hat{\Gamma}_{mn}$ which correspond to the ``longitudinal'' $(a,b)=(\parallel,\parallel)$, ``transverse'' $(a,b)=(\perp,\perp)$ and ``chiral'' $(a,b)=(\perp,\perp')$ channels relative to ${\bf r}_{ij} = |{\bf r}_{ij}|\hat{\bf z}$. This is required by symmetry. Since
\begin{eqnarray}
&& \Gamma^{ab}\hat{s}_i^a\hat{s}_j^b = \Gamma^{\perp\perp}(\hat{s}_i^x\hat{s}_j^x+\hat{s}_i^y\hat{s}_j^y) + \Gamma^{\parallel\parallel}\hat{s}_i^z\hat{s}_j^z + \cdots \\
&& \quad = \Gamma^{\perp\perp}(\hat{s}_i^x\hat{s}_j^x+\hat{s}_i^y\hat{s}_j^y+\hat{s}_i^z\hat{s}_j^z) + (\Gamma^{\parallel\parallel}-\Gamma^{\perp\perp})\hat{s}_i^z\hat{s}_j^z + \cdots \nonumber
\end{eqnarray}
in the fixed basis, $\Gamma^{\perp\perp}$, corresponds to the Heisenberg interaction and $\Gamma^{\parallel\parallel}-\Gamma^{\perp\perp}$ corresponds to the Kitaev interaction. The ``chiral'' channel $\Gamma^{\perp\perp'}$ builds the Dzyaloshinskii-Moriya interaction. The following $\theta$ integration is also straight-forward, and the integral over $q$ can be carried out exactly as well at least in the ``longitudinal'' and ``transverse'' channels (although the final analytical expressions are somewhat complicated and not particularly insightful).

\begin{figure}[!t]
\subfigure[{}]{\includegraphics[width=3.3in]{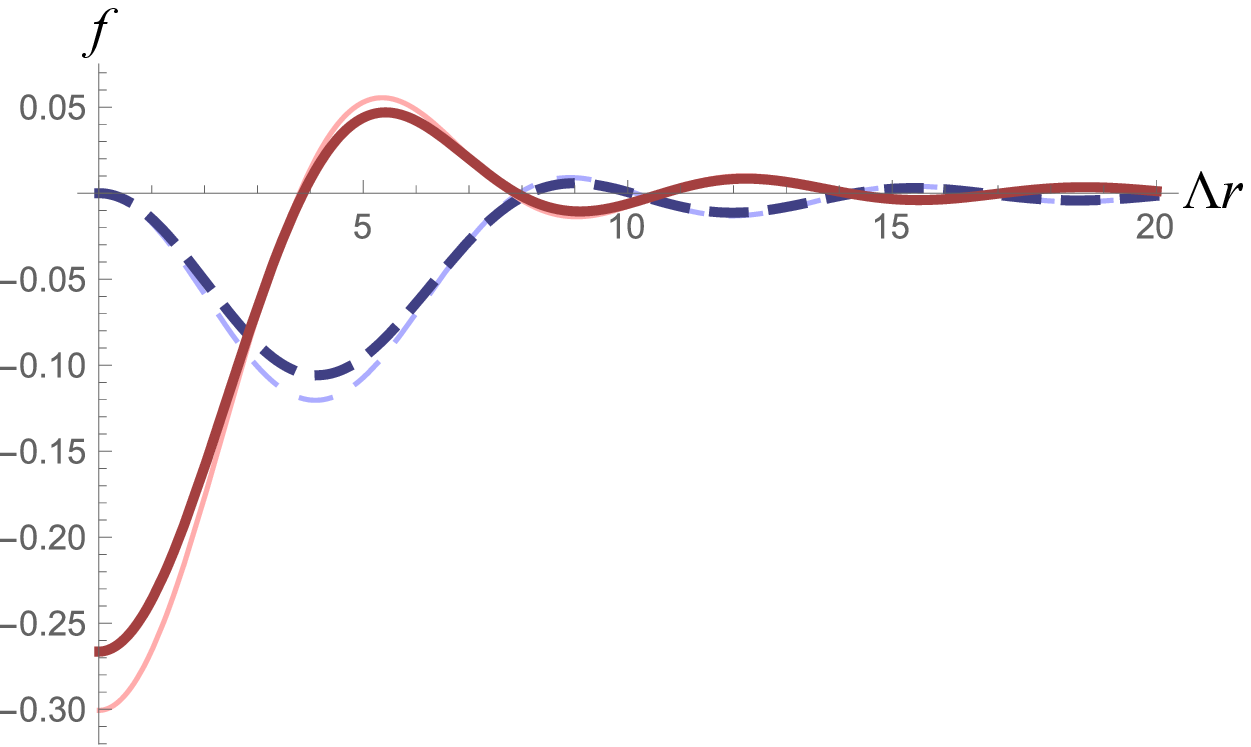}}
\subfigure[{}]{\includegraphics[width=3.3in]{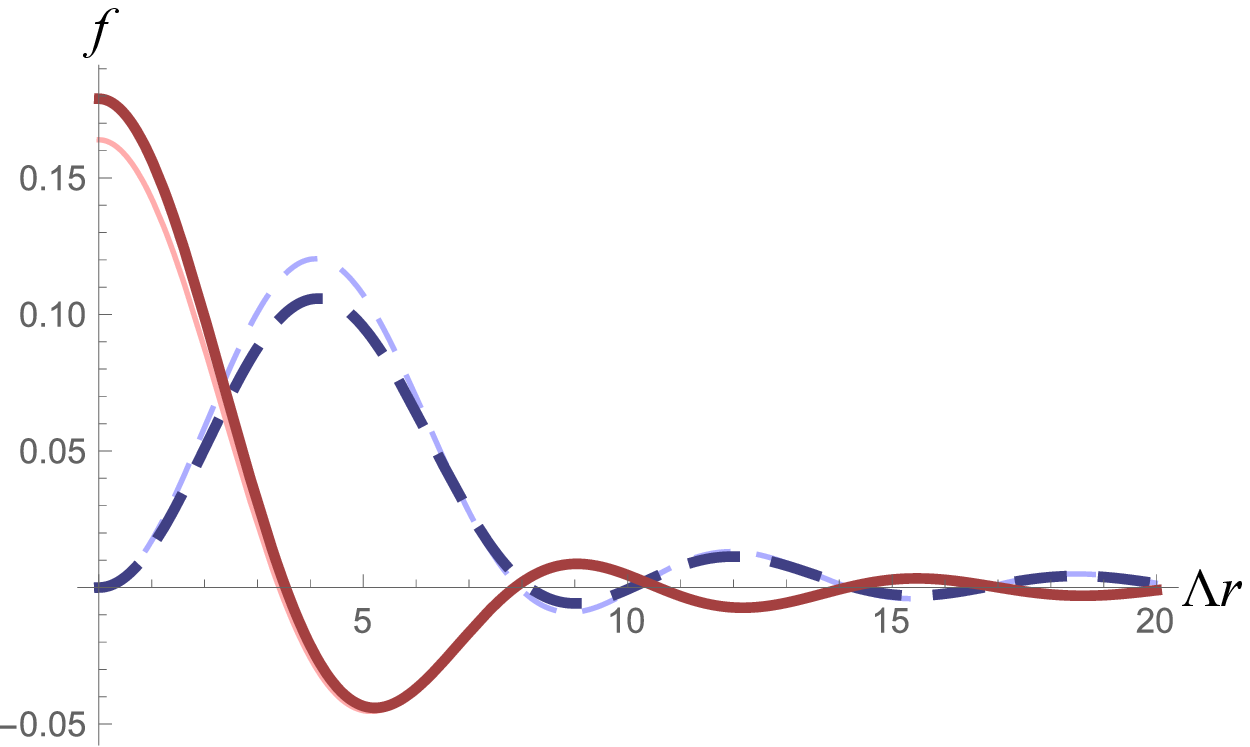}}
\subfigure[{}]{\includegraphics[width=3.3in]{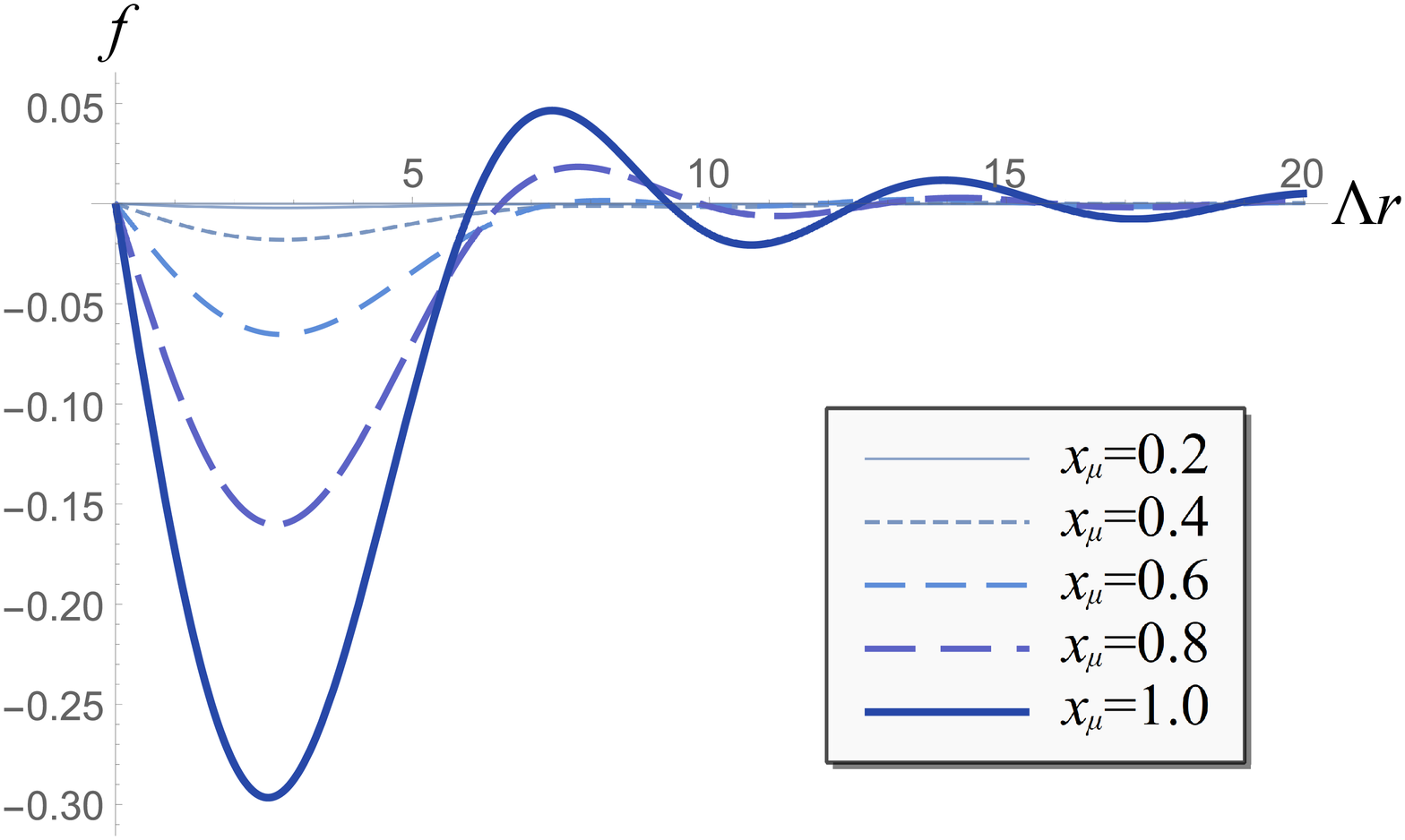}}
\caption{\label{fPlots}The plots of functions $f^{\textrm{H}}$, $f^{\textrm{K}}$ and $f^{\textrm{DM}}$ that capture the dependence of the two-spin interactions on the distance between the spins in the Hamiltonian (\ref{H2spin}). (a) and (b) correspond to the same-chirality and opposite-chirality channels respectively, showing the Heisenberg $f^{\textrm{H}}$ (red, solid) and Kitaev $f^{\textrm{K}}$ (blue, dashed) functions at $\mu=0$. The fainter thinner curves are obtained without the $q^2$ terms in (\ref{Gamma2e1}) and (\ref{Gamma2e2}) in order to demonstrate that zeroth-order terms in the expansions over $q$ make the decidedly dominant contributions. The amplitudes of these functions receive non-analytic corrections of the order of $|\mu|/v\Lambda$ when the chemical potential is away from the Weyl nodes. (c) shows the Dzyaloshinskii-Moriya function $f^{\textrm{DM}}$ for several values of $x_\mu = 2|\mu|/v\Lambda$; the trend of amplitude increase with $x_\mu$ is eventually halted near the cut-off. All functions have an envelope $\sim (\Lambda r)^{-2}$ at large $r$.}
\end{figure}

Ultimately, we represent the two-spin interactions
\begin{equation}
H_{\textrm{eff}} = \sum_{m,n} (H_{\textrm{H},mn} + H_{\textrm{K},mn} + H_{\textrm{DM},mn}) + \cdots
\end{equation}
as combinations of Heisenberg (H), Kitaev (K) and Dzyaloshinskii-Moriya (DM) couplings between the rectified spins (\ref{RectSpin}) in each node-pair channel:
\begin{eqnarray}\label{H2spin}
H_{\textrm{H},mn}^{\phantom{x}} &=& \frac{a^6 J_{\textrm{K}}^2 \Lambda^5}{(2\pi)^4 v} \sum_{ij} f_{mn}^{\textrm{H}}(\Lambda |{\bf r}_{ij}|) \; \hat{\bf s}_i^{\phantom{x}} \hat{\bf s}_j^{*} \\
H_{\textrm{K},mn}^{\phantom{x}} &=& \frac{a^6 J_{\textrm{K}}^2 \Lambda^5}{(2\pi)^4 v} \sum_{ij} f_{mn}^{\textrm{K}}(\Lambda |{\bf r}_{ij}|)\; (\hat{\bf s}_i^{\phantom{x}} \hat{\bf r}_{ij}^{\phantom{x}}) (\hat{\bf s}_j^{*} \hat{\bf r}_{ij}^{\phantom{x}}) \nonumber \\
H_{\textrm{DM},mn}^{\phantom{x}} &=& \frac{\chi_{m}+\chi_{n}}{2}\,\frac{\textrm{sign}(\mu)}{8}\,\frac{a^{6}J_{\textrm{K}}^{2}\Lambda^{5}}{(2\pi)^{4}v} \nonumber \\
&& \times \sum_{ij} f^{\textrm{DM}}(\Lambda |{\bf r}_{ij}|)\; \hat{\bf r}_{ij}^{\phantom{x}} (\hat{\bf s}_i^{\phantom{x}}\!\!\times \hat{\bf s}_j^{*}) \nonumber \ .
\end{eqnarray}
Here, $\hat{\bf r}_{ij}$ is the unit-vector along ${\bf r}_{ij}={\bf r}_i-{\bf r}_j$, and the dimensionless functions $f^{\textrm{H}}$, $f^{\textrm{K}}$ and $f^{\textrm{DM}}$ are plotted in Fig.\ref{fPlots}. By grouping together the $(i,j)$ and ($j,i$) terms in the sums, we immediately obtain manifestly Hermitian Hamiltonians in terms of the microscopic spins:\begin{equation}
\sum_{ij}f_{ij}^{ab} \hat{s}_i^a \hat{s}_j^{b*} = \sum_{ij}f_{ij}^{ab}
  \cos\Bigl(({\bf Q}_n-{\bf Q}_m)({\bf r}_i-{\bf r}_j)\Bigr) \hat{n}_i^a \hat{n}_j^b \ .
\end{equation}
This makes it apparent that $\hat{\bf n}_i$ want to make sign-changing oscillations at the wavevector(s) ${\bf Q}_m-{\bf Q_n}$, which can even be collinear as in NdAlSi \cite{Gaudet2021}.

If the neighboring local moments are separated by the lattice constant $a$, then their induced interactions have an extended range of the order of 5-10 $(\Lambda a)^{-1}$ lattice sites (note that $\Lambda a<1$). The Heisenberg interaction is peaked at short distances, being ferromagnetic (for the rectified spins $\hat{\bf s}_i$) in the same-chirality channels and antiferromagnetic in the opposite-chirality channels. A ferromagnetic coupling will simply realize a microscopic $\hat{\bf n}_i$ spin texture modulated at the inter-node wavevector(s) ${\bf Q}_n - {\bf Q}_m$. Every Weyl node contributes one same-chirality channel without a modulation from the intra-node electron scattering, and it is naively expected that these channels are most influential in the ultimate spin texture. At the same time, the induced Kitaev interactions are peaked at finite distances between the spins, and have the same ferromagnetic/antiferromagnetic character (for the rectified spins) as the Heisenberg coupling in the same channel. The resulting spin orientations are preferentially along the lattice bonds at least in the ferromagnetic same-chirality channels. The opposite-chirality channels with extended-range antiferromagnetic couplings are frustrated and may end up favoring spin orientations away from the lattice bonds.

Perhaps the most interesting induced interactions in this system are Dzyaloshinskii-Moriya (DM). They vanish at the relativistic point $\mu=0$ and become large when the Fermi pockets on the Weyl nodes have significant size. The sign of the DM interaction is different for electron and hole pockets, and only the same-chirality channels (including the intra-node electron scattering) contribute to it. The DM interaction is peaked for the spins separated by about $2-5 (\Lambda a)^{-1}$ lattice sites, and its maximum strength is not much below that of the Heisenberg and Kitaev interactions. The effect of such DM couplings on spin textures remains to be studied, but in general one expects a tendency to twist the spins into incommensurate ``spiral'' patterns, possibly developing hedgehogs \cite{Nikolic2019b}.

\section{Three-spin interactions}\label{sec3spin}

The coupling of itinerant electrons to local moments can generate the chiral spin interaction $\epsilon_{ijk}^{\phantom{x}}\epsilon^{abc}\hat{n}_{i}^{a}\hat{n}_{j}^{b}\hat{n}_{k}^{c}$ on triplets of lattice sites $i,j,k$. This is the only three-spin interaction allowed by the spin-rotation symmetry, but it requires a broken time-reversal (TR) symmetry. One way to break TR is spontaneously through the magnetic state of local moments. In such a state, the local moments impose an effective Zeeman field on itinerant electrons, which reorganizes the electron spectrum in a manner that enables the emergence of a chiral spin interaction. Capturing this effect requires at least fourth order of perturbation theory because at least four local spins are involved: one in the Zeeman coupling and three in the chiral interaction. Then, the ensuing chiral interaction is driven by the ``magnetic'' flux of the spin-orbit SU(2) gauge field through $ijk$ lattice plaquettes as in the Hubbard model \cite{Nikolic2019b}. We will not pursue here this complicated and perturbatively weak effect.

The presence of an external magnetic field ${\bf B}$ violates the TR symmetry and induces a chiral interaction at the third order of perturbation theory. We will derive the chiral interaction here by calculating the three-leg diagram in Fig.\ref{Bubbles}(b)
\begin{eqnarray}\label{Gamma3}
&& \Gamma^{abc}({\bf q}_{21},{\bf q}_{13})\equiv\Gamma^{abc}(q_{1},q_{2},q_{3}) \\
&& ~ = i\frac{J_{K}^{3}}{3}\!\int\!\!\frac{d^{4}k}{(2\pi)^{4}}\,\textrm{tr}\Bigl\lbrack G(k+q_{1})\sigma^{a}G(k+q_{2})\sigma^{b}G(k+q_{3})\sigma^{c}\Bigr\rbrack \nonumber
\end{eqnarray}
whose external legs correspond to $\hat{n}^{a}(q_{2}-q_{1})$, $\hat{n}^{b}(q_{3}-q_{2})$, $\hat{n}^{c}(q_{1}-q_{3})$. Writing the incoming 4-momenta of local moments as differences between $q_{1},q_{2},q_{3}$ achieves a convenient formal symmetry at the expense of a small redundancy: adding the same vector to all $q_i$ is inconsequential, so we will also work with ${\bf q}_{21} = {\bf q}_2-{\bf q}_1$, ${\bf q}_{13} = {\bf q}_1-{\bf q}_3$ as independent vectors when needed. We will include only the Zeeman effect of the external magnetic field ${\bf B}$ in the electron Hamiltonian
\begin{equation}\label{WeylHB}
H_{n}({\bf k})=(v\chi_{n}{\bf k}-{\bf B})\boldsymbol{\sigma}-\mu
\end{equation}
and neglect the orbital effect which is more fragile in the presence of disorder.

Just as in the case of two-spin interactions, we need to work with small momentum displacements ${\bf q}, {\bf k}$ from the Weyl node wavevectors ${\bf Q}_n$ in order to use the Green's function (\ref{Green2}) with the low-energy Hamiltonian (\ref{WeylHB}). This amounts to selecting any three Weyl nodes (labeled $1,2,3$) and interpreting all wavevectors in (\ref{Gamma3}) as ``small'' displacements from ${\bf Q}_1$, ${\bf Q}_2$, ${\bf Q}_3$. The obtained Feynman diagrams $\Gamma_{123}^{abc}$ should be eventually summed over all node triplets to obtain the three-spin interaction coupling $J_{ijk}^{abc}$ in real space.

Using the electron Green's functions (\ref{Green2}), the trace in (\ref{Gamma3}) is found to be:
\begin{eqnarray}\label{Trace3}
&& \textrm{tr}\left\lbrack G_1^{\phantom{x}}\sigma^a G_2^{\phantom{x}}\sigma^b G_3^{\phantom{x}}\sigma^c \right\rbrack \equiv \mathcal{T}_{123}^{abc} = X_{123}^{abc} \,\prod_{s=\pm1} \\
&& ~~\times \frac{1}{\omega_1\!+\!\mu\!-\!s\bigl|v\chi_1{\bf k}_1-{\bf B}\bigr|\!+\!i0^+\textrm{sign}\left(s\bigl|v\chi_1{\bf k}_1-{\bf B}\bigr|\!-\!\mu\right)} \nonumber \\
&& ~~\times \frac{1}{\omega_2\!+\!\mu\!-\!s\bigl|v\chi_2{\bf k}_2-{\bf B}\bigr|\!+\!i0^+\textrm{sign}\left(s\bigl|v\chi_2{\bf k}_2-{\bf B}\bigr|\!-\!\mu\right)} \nonumber \\
&& ~~\times \frac{1}{\omega_3\!+\!\mu\!-\!s\bigl|v\chi_3{\bf k}_3-{\bf B}\bigr|\!+\!i0^+\textrm{sign}\left(s\bigl|v\chi_3{\bf k}_3-{\bf B}\bigr|\!-\!\mu\right)} \nonumber
\end{eqnarray}
where $\omega_n = \omega + \Omega_n$ and ${\bf k}_n = {\bf k}+{\bf q}_n$ for $n=1,2,3$ are introduced to shorten the notation. Obtaining the factor $X_{123}^{abc}$ is straight-forward but tedious; $X_{123}^{abc}$ has many terms. We will dramatically simplify the analysis by discarding the irrelevant parts of $X_{123}^{abc}$ on the basis of symmetry. The general three-spin interaction
\begin{equation}
H_3 = \sum_{ijk} J_{ijk}^{abc} \hat{n}_i^a \hat{n}_j^b \hat{n}_k^c 
\end{equation}
is consistent with lattice rotation and spin SU(2) symmetries only if $J_{ijk}^{abc} = J \epsilon_{ijk}^{\phantom{x}} \epsilon^{abc}$. The spin-orbit coupling among the itinerant electrons promotes the global SU(2) symmetry into a gauge symmetry, by the virtue of admitting an effective SU(2) gauge field $A_{ij}^a$ coupled to spin currents $j_{ij}^a \sim \epsilon^{abc} \hat{n}_i^b \hat{n}_j^c$ (the spin-vector ${\bf A}_{ij}$ on the lattice bond ${ij}$ points parallel to the bond orientation ${\bf r}_i - {\bf r}_j$ in a Weyl semimetal). However, the index structure of $A_{ij}^a$ admits only gauge-invariant scalar contributions in the makeup of the chiral spin interaction, so we still have the symmetry requirement 
$J_{ijk}^{abc} = J \epsilon_{ijk}^{\phantom{x}} \epsilon^{abc}$. The kernel $J$ is sensitive only to a subset of the trace (\ref{Trace3}) terms $\mathcal{T}_{123}^{abc}$. Let us write (\ref{Jeff1}) explicitly
\begin{equation}
J_{ijk}^{abc} = J\epsilon_{ijk}^{\phantom{x}}\epsilon^{abc} \propto \hat{\mathcal{F}}_{ijk}^{123}\mathcal{T}_{123}^{abc}
\end{equation}
using the integration operator
\begin{eqnarray}\label{IntOp}
\hat{\mathcal{F}}_{ijk}^{123} &\equiv&
 3^3\!\! \int\!\!\frac{d^{3}q_{1}}{(2\pi)^{3}}\frac{d^{3}q_{2}}{(2\pi)^{3}}\frac{d^{3}q_{3}}{(2\pi)^{3}}\frac{d^{4}k}{(2\pi)^{4}}(2\pi)^{4}\delta^{3}({\bf q}_{1}\!+\!{\bf q}_{2}\!+\!{\bf q}_{3}) \nonumber \\
&& \times e^{i({\bf q}_{2}-{\bf q}_{1}){\bf r}_{i}}e^{i({\bf q}_{3}-{\bf q}_{2}){\bf r}_{j}}e^{i({\bf q}_{1}-{\bf q}_{3}){\bf r}_{k}}
\end{eqnarray}
(the factor $3^3$ is the Jacobian of the transformation from the momentum and frequency variables in (\ref{Jeff1}) to the symmetrized ones used here). This immediately reveals that only the components behaving as $\mathcal{T}_{123}^{abc}\propto\epsilon^{abc}$ contribute to $J$. The following argument will establish that the relevant terms also satisfy $\mathcal{T}_{lmn}^{abc}\propto\epsilon_{lmn}^{\phantom{x}}$. An innocuous cyclic permutation of the lattice site labels $ijk$ can be undone in $\hat{\mathcal{F}}_{ijk}^{123}$ by a cyclic exchange of $q_n$ integration variables, at the expense of inducing a cyclic permutation of the $123$ indices in $\mathcal{T}_{123}^{abc}$. Hence, the relevant $\mathcal{T}_{lmn}^{abc}$ terms are invariant under cyclic permutations of $lmn$. An order-changing permutation of $ijk$ changes the sign of the chiral coupling $J_{ijk}^{abc}$ and requires more care. It can be still compensated in $\hat{\mathcal{F}}_{ijk}^{123}$ by an exchange of two $q_n$ integration variables, but this flips the signs $e^{i(q_m-q_n)} \to e^{-i(q_m-q_n)}$ in all three exponential factors of (\ref{IntOp}). The remedy is to again change the integration variables, as $q_n\to-q_n, k_n\to-k_n$. Now, the original $\hat{\mathcal{F}}_{ijk}^{123}$ is restored, but the $\mathcal{T}_{lmn}^{abc}$ factor takes all frequency and momentum variables with reversed signs in addition to having its $lmn$ indices in the altered order. An inspection of the Green's functions reveals that the trace has the property:
\begin{equation}\label{Trace3b}
\mathcal{T}_{123}^{abc}(-q_n,-k_n;\mu,B) \to -\mathcal{T}_{123}^{abc}(q_n,k_n;-\mu,-B) \ .
\end{equation}
Coincidentally, charge conjugation transforms the Weyl Hamiltonian (\ref{WeylHB}) by $\chi_{n}\to-\chi_{n}$, $\mu\to-\mu$ ($B$ does not change sign). Since the effective local moment Hamiltonian is charge-neutral, the chiral coupling must behave as $J(\mu,B)=J(-\mu,B)$. We also have $J(\mu,B)=-J(\mu,-B)$ under time reversal. This allows us to absorb the last remaining sign changes of $\mu$ and $B$ in (\ref{Trace3b}) and conclude that the relevant parts of the trace (\ref{Trace3}) behave as $\mathcal{T}_{lmn}^{abc} \propto \epsilon_{lmn}^{\phantom{x}}$. Note that the product of denominators in (\ref{Trace3}) is invariant under all permutations of the $123$ indices, so we only need to calculate the parts of $X_{123}^{abc}$ that transform according to
\begin{equation}\label{Trace3bb}
X_{lmn}^{abc} = X(q_1^{\phantom{x}},q_2^{\phantom{x}},q_3^{\phantom{x}},k)\, \epsilon_{lmn}^{\phantom{x}} \epsilon^{abc} \ .
\end{equation}
Using again the short-hand notation $k_n = k+q_n$, the calculation of (\ref{Trace3}) yields:
\begin{eqnarray}\label{Trace3c}
&& X = \frac{1}{(3!)^2}\epsilon_{lmn}^{\phantom{x}} \epsilon^{abc} X_{lmn}^{abc} \\
&& \quad = \frac{1}{3!}\epsilon^{abc} \Bigl\lbrack -4v^{3}\chi_{1}^{\phantom{x}}\chi_{2}^{\phantom{x}}\chi_{3}^{\phantom{x}}k_{1}^{a}k_{2}^{b}k_{3}^{c} \nonumber \\
&& \qquad +4v^{2}B^{a}(\chi_{2}^{\phantom{x}}\chi_{3}^{\phantom{x}}k_{2}^{b}k_{3}^{c}+\chi_{3}^{\phantom{x}}\chi_{1}^{\phantom{x}}k_{3}^{b}k_{1}^{c}+\chi_{1}^{\phantom{x}}\chi_{2}^{\phantom{x}}k_{1}^{b}k_{2}^{c}) \Bigr\rbrack \ . \nonumber
\end{eqnarray}
We will greatly benefit from extracting the dependence of this and other expressions on the combinations of external wavevectors which are invariant under cyclic permutations:
\begin{eqnarray}\label{Q1}
\triangle{\bf q} &\equiv& {\bf q}_{2}\times{\bf q}_{1}+{\bf q}_{3}\times{\bf q}_{2}+{\bf q}_{1}\times{\bf q}_{3} \\
\triangle{\bf q}' &\equiv& \chi_{1}({\bf q}_{3}\times{\bf q}_{2})+\chi_{2}({\bf q}_{1}\times{\bf q}_{3})+\chi_{3}({\bf q}_{2}\times{\bf q}_{1}) \nonumber \\
\widetilde{{\bf q}} &\equiv& \chi_{1}({\bf q}_{3}-{\bf q}_{2})+\chi_{2}({\bf q}_{1}-{\bf q}_{3})+\chi_{3}({\bf q}_{2}-{\bf q}_{1}) \ . \nonumber
\end{eqnarray}
Defining ${\bf q}_{mn} = {\bf q}_m-{\bf q}_n$ and recalling ${\bf q}_1+{\bf q}_2+{\bf q}_3=0$, we also have ${\bf q}_1({\bf q}_2\times{\bf q}_3)=0$ and:
\begin{eqnarray}\label{Q2}
\triangle{\bf q}' &=& \frac{\chi_{1}+\chi_{2}+\chi_{3}}{3}\Delta{\bf q} \\
\triangle{\bf q} &=& {\bf q}_{21}\times{\bf q}_{13}={\bf q}_{32}\times{\bf q}_{21}={\bf q}_{13}\times{\bf q}_{32} \nonumber
\end{eqnarray}
Then, expressing (\ref{Trace3c}) in terms of the original integration variable ${\bf k}$ gives us
\begin{eqnarray}\label{Trace3d}
X &=& \frac{4\chi_1\chi_2\chi_3 v^2}{3!} \biggl\lbrack v\,{\bf k}\,\Delta{\bf q} \\
&& \qquad\qquad\quad -\frac{\chi_{1}+\chi_{2}+\chi_{3}}{3}{\bf B}\Delta{\bf q}+({\bf B}\times{\bf k})\widetilde{{\bf q}}\biggr\rbrack \nonumber
\end{eqnarray}

Having the trace $\mathcal{T}_{lmn}^{abc} \propto \epsilon_{lmn}^{\phantom{x}} \epsilon^{abc} X$, we can proceed with the derivation of (\ref{Gamma3}) specializing to a particular triplet $123$ of Weyl nodes:
\begin{widetext}
\begin{eqnarray}\label{Gamma3b}
&& \Gamma_{123}^{abc} = i\frac{J_{K}^{3}}{3}\! \int\!\frac{d^{4}k}{(2\pi)^{4}} \mathcal{T}_{123}^{abc} 
 = -\frac{2\chi_{1}\chi_{2}\chi_{3}J_{K}^{3}v^{2}}{3\cdot3!}\epsilon^{abc}\!\int\!\frac{d^{3}k}{(2\pi)^{3}}\left\lbrack v\,{\bf k}\,\Delta{\bf q}-\frac{\chi_{1}+\chi_{2}+\chi_{3}}{3}{\bf B}\Delta{\bf q}+({\bf B}\times{\bf k})\widetilde{{\bf q}}\right\rbrack \\
&& ~~\times\left\lbrack \frac{\theta(\mu\!-\!|v\chi_{1}{\bf k}_{1}\!-\!{\bf B}|)-\theta(\mu\!+\!|v\chi_{1}{\bf k}_{1}\!-\!{\bf B}|)}{|v\chi_{1}{\bf k}_{1}\!-\!{\bf B}|}\,\frac{1}{|v\chi_{1}{\bf k}_{1}\!-\!{\bf B}|^{2}-|v\chi_{2}{\bf k}_{2}\!-\!{\bf B}|^{2}}\,\frac{1}{|v\chi_{1}{\bf k}_{1}\!-\!{\bf B}|^{2}-|v\chi_{3}{\bf k}_{3}\!-\!{\bf B}|^{2}}+\textrm{cyclic}_{123}\right\rbrack \nonumber \ .
\end{eqnarray}
\end{widetext}
We substituted the trace (\ref{Trace3}), (\ref{Trace3bb}), (\ref{Trace3d}) and integrated out the loop frequency $\omega$. The cyclic index permutations in the second square bracket produce three terms; we will change integration variables ${\bf k}\to{\bf k}-{\bf q}_n+\chi_n{\bf B}/v$ in each term $n=1,2,3$ and use the properties
\begin{equation}
{\bf q}_{n}\Delta{\bf q}=0\quad,\quad{\bf q}_{n}\times\widetilde{{\bf q}}=\left(\chi_{n}-\frac{\chi_{1}+\chi_{2}+\chi_{3}}{3}\right)\Delta{\bf q}
\end{equation}
that stem from (\ref{Q1}) and (\ref{Q2}). This yields:
\begin{widetext}
\begin{eqnarray}\label{Gamma3c}
&& \Gamma_{123}^{abc} = -\frac{2\chi_{1}\chi_{2}\chi_{3}J_{K}^{3}v^{2}}{3\cdot3!}\epsilon^{abc}\!\int\!\frac{d^{3}k}{(2\pi)^{3}} \\
&& \qquad\times \biggl\lbrack \frac{\theta(\mu\!-\!v|{\bf k}|)-\theta(\mu\!+\!v|{\bf k}|)}{v|{\bf k}|}\,\frac{{\bf k}\,(v\,\Delta{\bf q}-{\bf B}\times\widetilde{{\bf q}})}{\left(v^{2}|{\bf k}|^{2}-|v({\bf k}\!+\!{\bf q}_{21})+(\chi_{1}\!-\!\chi_{2}){\bf B}|^{2}\right)\left(v^{2}|{\bf k}|^{2}-|v({\bf k}\!-\!{\bf q}_{13})+(\chi_{1}\!-\!\chi_{3}){\bf B}|^{2}\right)}+\textrm{cyclic}_{123}\biggr\rbrack \nonumber
\end{eqnarray}
\end{widetext}
We will integrate out the wavevector ${\bf k}$ in polar coordinates. One can easily show using (\ref{Q1}) that
\begin{eqnarray}
\Bigl(v{\bf q}_{21}+(\chi_{1}-\chi_{2}){\bf B}\Bigr)(v\triangle{\bf q}-{\bf B}\times\widetilde{{\bf q}})&=&0 \nonumber \\
\Bigl(-v{\bf q}_{13}+(\chi_{1}-\chi_{3}){\bf B}\Bigr)(v\triangle{\bf q}-{\bf B}\times\widetilde{{\bf q}})&=&0 \nonumber \ .
\end{eqnarray}
If we decompose the vector ${\bf k}={\bf k}_\parallel + {\bf k}_\perp$ into the component ${\bf k}_{\parallel}=k_{\parallel}\hat{{\bf l}}$ parallel to $v\,\Delta{\bf q}-{\bf B}\times\widetilde{{\bf q}}=|v\,\Delta{\bf q}-{\bf B}\times\widetilde{{\bf q}}|\hat{{\bf l}}$ and the component ${\bf k}_{\perp}$ perpendicular to it, then ${\bf k}_{\parallel}$ is perpendicular to both $v{\bf q}_{21}+(\chi_{1}-\chi_{2}){\bf B}$ and $-v{\bf q}_{13}+(\chi_{1}-\chi_{3}){\bf B}$. Hence:
\begin{widetext}
\begin{eqnarray}\label{Gamma3d}
&&\Gamma_{123}^{abc} = -\frac{2\chi_{1}\chi_{2}\chi_{3}J_{K}^{3}v^{2}}{3\cdot3!}\epsilon^{abc}\!\int\!\frac{d^{2}k_\perp}{(2\pi)^{2}}\int\limits_{k_-}^{k_+}\frac{dk_\parallel}{2\pi} \Biggl\lbrack \frac{\theta\left(\mu\!-\!v\sqrt{k_{\parallel}^{2}\!+\!k_{\perp}^{2}}\right)-\theta\left(\mu\!+\!v\sqrt{k_{\parallel}^{2}\!+\!k_{\perp}^{2}}\right)}{v\sqrt{k_{\parallel}^{2}+k_{\perp}^{2}}} \\
&& \qquad\times\frac{k_{\parallel}\,|v\,\Delta{\bf q}-{\bf B}\times\widetilde{{\bf q}}|}{\left(v^{2}k_{\perp}^{2}-|v({\bf k}_{\perp}+{\bf q}_{21})+(\chi_{1}-\chi_{2}){\bf B}|^{2}\right)\left(v^{2}k_{\perp}^{2}-|v({\bf k}_{\perp}-{\bf q}_{13})+(\chi_{1}-\chi_{3}){\bf B}|^{2}\right)}+\textrm{cyclic}_{123}\Biggr\rbrack \nonumber
\end{eqnarray}
\end{widetext}
Naively, the integral over $k_\parallel$ should vanish by being the integral of an odd function in a symmetric interval. However, this integral needs to be cut-off by $\Lambda$ and the integration interval is symmetric only \emph{before} shifting ${\bf k}$ by the amount proportional to the external field ${\bf B}$. Limiting the original unshifted ${\bf k}$ by $\Lambda$ corresponds to
\begin{equation}
k_\pm = \pm\Lambda-({\bf q}_{i}-\chi_{i}{\bf B}/v)\hat{{\bf l}} = \pm\Lambda+({\bf B}\Delta{\bf q})\,\frac{\chi_{1}+\chi_{2}+\chi_{3}}{3|v\,\Delta{\bf q}-{\bf B}\times\widetilde{{\bf q}}|} \nonumber \ ,
\end{equation}
and the $k_\parallel$ integral in (\ref{Gamma3d}) receives contribution only from one of its boundary regions whose extent is proportional to ${\bf B}\Delta{\bf q}$. Since $k_\parallel$ is large in that region, we can take $k_\parallel \approx \pm\Lambda$ everywhere in the integral and neglect $k_\perp$, etc. next to it, thus keeping only the terms with the leading power of $\Lambda$. Interestingly, this leading power is $\Lambda^0$ since a diverging $k_\parallel$ cancels out in (\ref{Gamma3d}). Hence, integrating out $k_\parallel$ yields:
\begin{eqnarray}\label{Gamma3e}
&& \Gamma_{123}^{abc} = \frac{2J_{K}^{3}}{3\pi\cdot3!\,v^{3}}\,\frac{\chi_{1}\chi_{2}\chi_{3}(\chi_{1}+\chi_{2}+\chi_{3})}{3}\,({\bf B}\Delta{\bf q})\,\epsilon^{abc} \nonumber \\
&& \quad \times\int\frac{d^{2}k_{\perp}}{(2\pi)^{2}} \biggl\lbrack \frac{1}{\left(k_{\perp}^{2}-|{\bf k}_{\perp}+{\bf q}_{21}|^{2}\right)\left(k_{\perp}^{2}-|{\bf k}_{\perp}-{\bf q}_{13}|^{2}\right)} \nonumber \\
&& \qquad\qquad\qquad +\textrm{cyclic}_{123}\biggr\rbrack + \cdots
\end{eqnarray}
with lower powers of $\Lambda$ and higher orders of ${\bf B}$ (the dots) neglected. Let us label the remaining cyclically permuted integrals over ${\bf k}_\perp$ by $I_1+I_2+I_3$. We can integrate out ${\bf k}_{\perp}=(k_\perp,\theta)$ using polar coordinates. In $I_1$ for example, align the $x$-axis with ${\bf q}_{21}$ and assume that ${\bf q}_{13}$ makes the angle $\phi$ with it. Then, writing $q_{21}=|{\bf q}_{21}|$ and $q_{13}=|{\bf q}_{13}|$ we get:
\begin{eqnarray}\label{I1}
I_1 &=& -\frac{1}{(2\pi)^{2}q_{21}q_{13}}\int\limits_0^\Lambda dk_{\perp}\,\int\limits_{0}^{2\pi}d\theta \\
&& \times \frac{k_{\perp}}{\left(2k_{\perp}\cos\theta+q_{21}\right)\left(2k_{\perp}\cos(\theta-\phi)-q_{13}\right)} \nonumber \ .
\end{eqnarray}
The $\theta$ integral is of the form:
\begin{eqnarray}
&& \mathbb{P}\int\limits_{0}^{2\pi}d\theta\,\frac{1}{\Bigl(\cos\theta+a\Bigr)\Bigl(\cos(\theta-\phi)-b\Bigr)} \\
&& \qquad = -4\pi\frac{(a\cos\phi+b)\frac{\theta(a-1)}{\sqrt{a^{2}-1}}+(b\cos\phi+a)\frac{\theta(b-1)}{\sqrt{b^{2}-1}}}{2a^{2}+4ab\cos\phi+2b^{2}+\cos(2\phi)-1} \nonumber
\end{eqnarray}
and obtains using the Cauchy's residue theorem. We keep only its principal part because we are not interested in dissipation. Substituting in (\ref{I1}) and changing the integration variable into $\xi = 4k_{\perp}^{2}$ gives us:
\begin{eqnarray}
&& I_1 = \frac{\pi}{4(2\pi)^{2}q_{21}q_{13}} \\
&& ~ \times \Biggl\lbrace
\int\limits_{0}^{q_{21}^{2}}\!d\xi\,\frac{({\bf q}_{21}\!+\!{\bf q}_{13}){\bf q}_{13}}{q_{13}\sqrt{q_{21}^{2}\!-\!\xi}}\,\frac{1}{({\bf q}_{21}\!+\!{\bf q}_{13})^{2}+\xi \left(\frac{{\bf q}_{21}{\bf q}_{13}}{q_{21}q_{13}}\right)^{2}\!\!-\xi } \nonumber \\
&& \quad +\int\limits _{0}^{q_{13}^{2}}\!d\xi\,\frac{({\bf q}_{21}\!+\!{\bf q}_{13}){\bf q}_{21}}{q_{21}\sqrt{q_{13}^{2}\!-\!\xi}}\,\frac{1}{({\bf q}_{21}\!+\!{\bf q}_{13})^{2}+\xi \left(\frac{{\bf q}_{21}{\bf q}_{13}}{q_{21}q_{13}}\right)^{2}-\xi} \Biggr\rbrace \nonumber
\end{eqnarray}
The final $\xi$ integration is straight-forward. Using again the definition and properties (\ref{Q1}), (\ref{Q2}) of $\triangle{\bf q}$ we finally obtain
\begin{equation}
I_1+I_2+I_3 = -\frac{\pi W}{(2\pi)^{2}|\Delta{\bf q}|} \ ,
\end{equation}
where
\begin{equation}
W = \sum_{n=1}^3 \arctan\Bigl(\tan(\phi_n)\Bigr) \ ,
\end{equation}
and
\begin{equation}
\tan(\phi_1) = \frac{|{\bf q}_{21}\times{\bf q}_{13}|}{{\bf q}_{21}{\bf q}_{13}} \quad,\quad \dots
\end{equation}
involve the angles $\phi_1, \phi_2, \phi_3 \in (0,\pi)$ between the pairs of vectors $({\bf q}_{21},{\bf q}_{13}), ({\bf q}_{32},{\bf q}_{21}), ({\bf q}_{13},{\bf q}_{32})$ respectively. Since ${\bf q}_{21}+{\bf q}_{32}+{\bf q}_{13}=0$, these vectors lie in the same plane and form a triangle whose exterior angles are $\phi_1,\phi_2,\phi_3$. Note that $W$ is not simply the sum of $\phi_n$ due to
\begin{equation}
\arctan\Bigl(\tan(\phi_n\Bigr)=\begin{cases}
\phi_n & ,\quad\phi_n<\frac{\pi}{2}\\
\phi_n-\pi & ,\quad\phi_n>\frac{\pi}{2}
\end{cases} \ .
\end{equation}
Instead, $W=-\pi$ if all three angles $\phi_n$ are obtuse and $W=0$ otherwise. In other words, $W$ selects only acute triangles made by ${\bf q}_{21}, {\bf q}_{32}, {\bf q}_{13}$. Combining these conclusions with (\ref{Gamma3e}), we have:
\begin{equation}\label{Gamma3f}
\Gamma_{123}^{abc} = \frac{\epsilon^{abc}J_{K}^{3}}{3\cdot3!(2\pi)\,v^{3}}\,\frac{\chi_{1}\chi_{2}\chi_{3}(\chi_{1}\!+\!\chi_{2}\!+\!\chi_{3})}{3}\,\frac{{\bf B}\Delta{\bf q}}{|\Delta{\bf q}|}\,W_{123}^{\phantom{x}}
\end{equation}
where
\begin{equation}\label{AcuteFilter}
W_{123} = \begin{cases}
1 & ,\quad{\bf q}_{21},{\bf q}_{32},{\bf q}_{13}\textrm{ make an acute triangle}\\
0 & ,\quad\textrm{otherwise} \ .
\end{cases}
\end{equation}

\subsection{Induced chiral interactions in real space}\label{sec3spinRS}

Based on (\ref{Gamma3f}), here we obtain the interaction coupling
\begin{eqnarray}\label{Jeff3}
J_{ijk}^{abc} &=& -a^{9} \sum_{lmn} e^{i{\bf Q}_{ml}{\bf r}_{i}}e^{i{\bf Q}_{nm}{\bf r}_{j}}e^{i{\bf Q}_{ln}{\bf r}_{k}} \\
&& \times \int\frac{d^{3}q_{ml}}{(2\pi)^{3}}\frac{d^{3}q_{ln}}{(2\pi)^{3}}\,\Gamma_{lmn}^{abc}({\bf q}_{ml},{\bf q}_{ln})\,e^{i{\bf q}_{ml}{\bf r}_{ij}}e^{-i{\bf q}_{ln}{\bf r}_{jk}} \nonumber \\
&=& -\frac{\epsilon^{abc}a^{9}J_{K}^{3}}{3\cdot3!(2\pi)\,v^{3}}\sum_{lmn}\frac{\chi_{l}\chi_{m}\chi_{n}(\chi_{l}+\chi_{m}+\chi_{n})}{3} \nonumber \\
&& \times e^{i{\bf Q}_{ml}{\bf r}_{i}}e^{i{\bf Q}_{nm}{\bf r}_{j}}e^{i{\bf Q}_{ln}{\bf r}_{k}}B^{p}\mathcal{J}_{lmn}^{p} \nonumber
\end{eqnarray}
in the real-space Hamiltonian
\begin{equation}\label{Heff3}
H_{\textrm{eff}} = \cdots+\sum_{ijk}J_{ijk}^{abc}\,\hat{n}_{i}^{a}\hat{n}_{j}^{b}\hat{n}_{k}^{c}+\cdots \ .
\end{equation}
We will first focus on a particular $(l,m,n)=(1,2,3)$ triplet of Weyl nodes and $ijk$ triplet of lattice sites. We wrote ${\bf Q}_{mn}={\bf Q}_m-{\bf Q}_n$ and ${\bf r}_{ij}={\bf r}_i-{\bf r}_j$ to shorten the notation. The challenge is to compute the integral
\begin{equation}
\mathcal{J}_{123}^{a}=\int\frac{d^{3}q_{21}}{(2\pi)^{3}}\frac{d^{3}q_{13}}{(2\pi)^{3}}\,\frac{\epsilon^{abc}q_{21}^{b}q_{13}^{c}}{|{\bf q}_{21}\times{\bf q}_{13}|}\,e^{i{\bf q}_{21}{\bf r}_{ij}}e^{-i{\bf q}_{13}{\bf r}_{jk}}W_{123}
\end{equation}
that stems from (\ref{Gamma3f}). Due to the unbiased sampling of the momentum space, the vector $\mathcal{J}^a_{123}$ takes rotational bias only from ${\bf r}_{ij}$ and ${\bf r}_{jk}$. Exchanging ${\bf r}_{ij}$ and ${\bf r}_{jk}$ can be compensated by $({\bf q}_{21},{\bf q}_{13})\to(-{\bf q}_{13},-{\bf q}_{21})$, which changes the sign of $\mathcal{J}^a_{123}$. The property $\mathcal{J}_{123}^{a}({\bf r}_{jk},{\bf r}_{ij})=-\mathcal{J}_{123}^{a}({\bf r}_{ij},{\bf r}_{jk})$ and its vector transformations under rotations imply:
\begin{equation}
\mathcal{J}_{123}^{a}({\bf r}_{ij},{\bf r}_{jk})=\epsilon^{abc}r_{ij}^{b}r_{jk}^{c}\:\mathcal{J}_{123}^{\phantom{x}}\left(|{\bf r}_{ij}^{\phantom{x}}|,|{\bf r}_{jk}^{\phantom{x}}|,{\bf r}_{ij}^{\phantom{x}}{\bf r}_{jk}^{\phantom{x}}\right) \ .
\end{equation}
Hence, finding $\mathcal{J}^a_{123}$ reduces to the calculation of the scalar
\begin{eqnarray}\label{Jeff3b}
&& \mathcal{J}_{123}^{\phantom{x}} = \frac{\epsilon^{abc}r_{ij}^{b}r_{jk}^{c}}{|{\bf r}_{ij}\times{\bf r}_{jk}|^{2}}\mathcal{J}_{123}^{a} = \frac{{\bf r}_{ij}\times{\bf r}_{jk}}{|{\bf r}_{ij}\times{\bf r}_{jk}|^2} \\
&& \quad \times \int\frac{d^{3}q_{21}}{(2\pi)^{3}}\frac{d^{3}q_{13}}{(2\pi)^{3}}\,\frac{{\bf q}_{21}\times{\bf q}_{13}}{|{\bf q}_{21}\times{\bf q}_{13}|}\,e^{i{\bf q}_{21}{\bf r}_{ij}}e^{-i{\bf q}_{13}{\bf r}_{jk}}W_{123} \ . \nonumber
\end{eqnarray}
The presence of the acute triangle filter $W_{123}$ given by (\ref{AcuteFilter}) complicates this integral very much.

Since the Weyl node cut-off momentum $\Lambda$ is smaller than the microscopic lattice cut-off $\pi/a$, the effect of spin interactions is mostly felt on the short separations ${\bf r}$ between the spins which satisfy $\Lambda|{\bf r}|\ll1$. These separations can still span multiple lattice constants $a$. We may expand the exponential factors of (\ref{Jeff3b}) to quadratic order in the limit $\Lambda|{\bf r}|\ll1$ and approximately obtain after some manipulations:
\begin{eqnarray}
\mathcal{J}_{123}^{\phantom{x}} &\approx& \int\frac{d^{3}q_{21}}{(2\pi)^{3}}\frac{d^{3}q_{13}}{(2\pi)^{3}}\,\frac{\Bigl\lbrack({\bf r}_{ij}\times{\bf r}_{jk})({\bf q}_{21}\times{\bf q}_{13})\Bigr\rbrack^{2}}{2|{\bf r}_{ij}\times{\bf r}_{jk}|^2|{\bf q}_{21}\times{\bf q}_{13}|}W_{123} \nonumber \\
&=& \frac{1}{2}\int\frac{d^{3}q_{21}}{(2\pi)^{3}}\frac{d^{3}q_{13}}{(2\pi)^{3}}\,\frac{\Bigl\lbrack\hat{{\bf z}}({\bf q}_{21}\times{\bf q}_{13})\Bigr\rbrack^{2}}{|{\bf q}_{21}\times{\bf q}_{13}|}W_{123} \ .
\end{eqnarray}
Here, $\hat{\bf z}$ is the unit-vector along ${\bf r}_{ij}\times{\bf r}_{jk}$, and all dependence on ${\bf r}_{ij}$ or ${\bf r}_{jk}$ has dropped out of the integral. Consequently, the integral is controlled by a single scale $\Lambda$ and it must evaluate to $\mathcal{J}_{123} \approx \lambda \Lambda^8$ given its units, where the dimensionless constant $\lambda>0$ is hard to calculate. We have
\begin{equation}
\mathcal{J}_{123}^a \xrightarrow{\Lambda|{\bf r}|\ll 1}  \lambda \Lambda^8\, \epsilon^{abc}r_{ij}^{b}r_{jk}^{c} \ ,
\end{equation}
so the chiral spin interaction initially grows with the distance between the spins.

The chiral interaction eventually looses strength as a power law of the separation between the spins when $\Lambda|{\bf r}|\gg 1$. We can estimate these attenuation powers of $|{\bf r}|^{-1}$ in $\mathcal{J}_{123}$, but it should be noted that they depend on the exact manner in which the cut-off is imposed in the momentum integral. The most unbiased cut-off is attained by using the spherical coordinates for momentum integration and limiting the momentum magnitude to $q<\Lambda$. The integral in (\ref{Jeff3b}) has the units of $\Lambda^6$, but its exponential factors introduce destructive interference when $|{\bf q}_{21}{\bf r}_{ij}|\gg1$ or $|{\bf q}_{13}{\bf r}_{jk}|\gg1$, so at least two power units ($\Lambda^2$) are cut-off and replaced with $|{\bf r}_{ij}|^{-1} |{\bf r}_{jk}|^{-1}$. This would be all in cylindrical coordinates, but the use of spherical coordinates cuts-off another factor of $\Lambda^2$ through the integration of its polar angles $\theta$. The rest of the integrand performs a projection and filtering which directly affect only the unit-less angle integrations. Hence, we naively expect $\mathcal{J}_{123} \approx \lambda'\Lambda^2 / (|{\bf r}_{ij}|^2 |{\bf r}_{jk}|^2 |{\bf r}_{ij}\!\times\!{\bf r}_{jk}|)$ and
\begin{equation}
\mathcal{J}_{123}^a \xrightarrow{\Lambda|{\bf r}|\gg 1}  \frac{\lambda' \Lambda^2}{|{\bf r}_{ij}|^2 |{\bf r}_{jk}|^2}\, \frac{\epsilon^{abc}r_{ij}^{b}r_{jk}^{c}}{|{\bf r}_{ij}\times{\bf r}_{jk}|} \ .
\end{equation}
Here, $\lambda'$ is a complicated dimensionless oscillatory function of $\Lambda|{\bf r}_{ij}|$ and $\Lambda|{\bf r}_{jk}|$ which we shall not attempt to determine. A careful calculation of (\ref{Jeff3b}) may still qualitatively correct this naive expectation, but not in ways that make the attenuation rate slower than $\Lambda^4 / (|{\bf r}_{ij}| |{\bf r}_{jk}|)$ or faster than $1 / (|{\bf r}_{ij}|^3 |{\bf r}_{jk}|^3)$.

In summary, once all triplets $123$ of Weyl nodes and lattice sites $ijk$ are summed up, we can express the chiral spin interaction in a manifestly symmetric form
\begin{equation}\label{Jeff3c}
J_{ijk}^{abc} = -\epsilon^{abc}\,\frac{a^{9}\Lambda^{6}J_{K}^{3}}{3\cdot3!(2\pi)\,v^{3}} \,\frac{{\bf B}\triangle{\bf r}_{ijk}}{|\triangle{\bf r}_{ijk}|}\,f_{ijk}
\end{equation}
where
\begin{equation}
\triangle{\bf r}_{ijk} = {\bf r}_{i}\times{\bf r}_{j}+{\bf r}_{j}\times{\bf r}_{k}+{\bf r}_{k}\times{\bf r}_i \nonumber
\end{equation}
and
\begin{eqnarray}
&& f_{ijk} = \sum_{lmn} \frac{\chi_{l}\!+\!\chi_{m}\!+\!\chi_{n}}{3\chi_{l}\chi_{m}\chi_{n}} e^{i({\bf Q}_{ml}{\bf r}_{i}+{\bf Q}_{nm}{\bf r}_{j}+{\bf Q}_{ln}{\bf r}_{k})} \\
&& \qquad\times \begin{cases}
\lambda\Lambda^{2}|\triangle{\bf r}_{ijk}| & ,\quad|{\bf r}_{ij}|,|{\bf r}_{jk}|,|{\bf r}_{ki}|\ll\frac{1}{\Lambda}\\
\frac{\lambda'}{3\Lambda^{4}}\,\frac{|{\bf r}_{ij}|^{2}+|{\bf r}_{jk}|^{2}+|{\bf r}_{ki}|^{2}}{|{\bf r}_{ij}|^{2}|{\bf r}_{jk}|^{2}|{\bf r}_{ki}|^{2}} & ,\quad|{\bf r}_{ij}|,|{\bf r}_{jk}|,|{\bf r}_{ki}|\gg\frac{1}{\Lambda} 
\end{cases} \nonumber
\end{eqnarray}
Note that
\begin{equation}
\frac{{\bf B}\triangle{\bf r}_{ijk}}{|\triangle{\bf r}_{ijk}|} = \hat{\boldsymbol{\zeta}} \, \epsilon_{ijk}
\end{equation}
brings the spatial Levi-Civita tensor to (\ref{Jeff3c}), where $\hat{\boldsymbol{\zeta}}$ is the unit-vector perpendicular to the triangle formed by the lattice sites $i,j,k$ but stripped of the triangle's orientation.

\section{Discussion and conclusions}\label{secDiscussion}

We analyzed the s-d model of local magnetic moments coupled to itinerant Weyl electrons, and derived the Weyl-electron-mediated interactions between the local spins. We obtained detailed forms of the Heisenberg, Kitaev and Dzyaloshinskii-Moriya (DM) interactions, and also characterized the chiral spin interaction in the presence of an external magnetic field. Due to the relativistic nature of Weyl electrons, the main length scale that controls these interactions is the momentum cut-off $\Lambda$ of the linear Weyl spectrum. The local Kondo coupling controls only the strength of interactions, while all other energy scales $\mathcal{E}$ (temperature, chemical potential, magnetic field, etc.) introduce nominally small corrections of the order of $\mathcal{E}/v\Lambda$, where $v$ is the Fermi velocity (slope of the Weyl electron's energy $\epsilon_{\bf k}$).


Making analytical progress was made possible by various idealizations. All Weyl nodes were assumed to be identical, isotropic, at the same energy and of type-I. Most of these simplifications are not qualitatively significant. Nodes living at different energies are most easily accommodated by associating different chemical potentials to different nodes. The consequences of this are expected to be small, since the chemical potential was found to act as a small perturbative parameter. Anisotropy is expected to introduce a related spatial anisotropy in the induced spin interactions. A more dramatic issue in the present perturbation theory is the implicit assumption that the Weyl electron spectrum is known and magnetically unbiased. This either neglects the effect of local moments on the electron dynamics, or presumes self-consistently that the considered Weyl spectrum is already (at least approximately) a result of the magnetic order that would arise from the interactions that we calculate. The full self-consistent problem of the mutual influence of local moments and Weyl electrons is hard and beyond the scope of this study. Nevertheless, some qualitative features of the self-consistent picture can be readily anticipated. A magnetic order of local moments will generally induce a compatible spin-density wave of itinerant electrons; its ferromagnetic part presents itself as an effective magnetic field to the local spin, which we included in the calculations. A reconstruction of the Weyl Fermi surfaces is implicitly included through the renormalization of the model parameters $v$, $\Lambda$, $\mu$ and changes in the number, chiralities and locations of the Weyl nodes. In specific cases, band-structure calculations can reveal how exactly the Weyl spectrum depends on the magnetic state \cite{Gaudet2021}.

Several problems are left for future studies. An accurate inclusion of anisotropy and other realistic features of the Weyl electron spectra most likely requires numerical calculations. Pursuing them will take full justification only from the desire to quantitatively explain the magnetism of concrete materials whose electronic spectrum is well understood. The effect of type-II Weyl nodes on the local moment magnetism remains a fundamental open problem worth analyzing. Beyond this, the prediction of chiral magnetic orders, phase diagrams, critical behaviors, etc. from the knowledge of interactions among the spins is at hand with the help of theoretical methods (mean field approximation, renormalization group) and numerical approaches (Monte Carlo) -- and obviously is an important component in the full understanding of magnetic topological materials.

This theory provides a plausible explanation of the magnetic order observed in the magnetic Weyl semimetal NdAlSi despite all complexities \cite{Gaudet2021}. Still, more research is needed to positively confirm the offered physical picture as opposed to some alternative scenario. For example, could the observed features of the NdAlSi magnetic order be shaped by conventional instead of Weyl Fermi pockets, assuming that their sizes and locations in momentum space are similar? Apart from trusting the band-structure calculations, the answer is in the currently unknown details. Conventional Fermi pockets are expected to embed their Fermi wavevector in the attenuated spatial oscillations of the induced spin coupling. In contrast, the relativistic Weyl electrons embed their much larger cut-off momentum, and the ensuing spin interactions have shorter range. All other things being equal, this would affect the spin stiffness and hence the spin dynamics. Furthermore, Weyl electrons are able to induce notable DM and chiral interactions, and thus encourage the twisting of the spin texture. Other ways to experimentally distinguish the influence of conventional electrons and Weyl nodes may come from the damping of spin waves, but this will be discussed in a forthcoming paper.

At this time, magnetic Weyl semimetals are rare, and the ones with only Weyl pockets in the Fermi surface are even more rare. For example, the well-known chiral magnets Mn$_3$Sn and Mn$_3$Ge \cite{Nakatsuji2015, Nakatsuji2016, Parkin2016} have significant conventional parts of the Fermi surface in addition to Weyl nodes \cite{Parkin2017, Felser2017b}. Perhaps an interesting direction to look for other candidate materials is Dirac semimetals. For example, the material YbMnBi$_2$ has only small conventional Fermi pockets coexisting with Dirac nodes, while hosting an antiferromagnetic order \cite{Borisenko2015, Chinotti2016, Armitage2017, Petrovic2016}. A Dirac node can be viewed as a coalescence of two opposite-chirality Weyl nodes at the same wavevector, so some insights of the theory developed here should apply. The predicted antiferromagnetic interactions induced by the opposite-chirality inter-node scattering now arise between the microscopic spins, which are identified with the ``rectified'' spins of the ensuing commensurate $\Delta{\bf Q}=0$ channel. This is naively consistent with the observed antiferromagnetic order, but there are many complications and open questions: (i) are the equal-chirality ferromagnetic channels stronger or weaker, (ii) is the extended range of antiferromagnetic interactions short enough to avoid frustration, (iii) what are the RKKY interactions due to the conventional Fermi pockets, (iv) are there intrinsic spin interactions between the moments, etc.?

Looking beyond the currently known materials, the induced spin interactions in Weyl semimetals have rich ingredients which by themselves are capable of stabilizing unconventional magnetic states. Kitaev interactions can lead to spin liquids as demonstrated with exactly solvable models \cite{Kitaev2006b}. The DM interactions generated by Weyl electrons are likely capable of producing magnetic states that host lattices or liquids of hedgehogs \cite{Nikolic2019b}. A hedgehog lattice melted by quantum or thermal fluctuations is a candidate for a chiral spin liquid state \cite{Nikolic2019} -- a three-dimensional analogue of the fractional quantum Hall liquid. This and similar future research will hopefully illuminate the path toward the realization of such states, which possess both a profound fundamental appeal and a potential for applications in quantum information processing.

\section{Acknowledgments}\label{secAck}

I am very grateful for insightful discussions and collaboration with Jonathan Gaudet and Collin Broholm. This research was supported at the Institute for Quantum Matter, an Energy Frontier Research Center funded by the U.S. Department of Energy, Office of Science, Basic Energy Sciences under Award No. DE-SC0019331.



%

\end{document}